\documentclass[
 preprint,
 amsmath,amssymb,
 aps,
]{revtex4-2}

\usepackage[utf8]{inputenc}
\usepackage[T1]{fontenc}
\usepackage{graphicx}
\usepackage{dcolumn}
\usepackage{bm}
\usepackage{hyperref}
\usepackage{braket,xcolor}
\usepackage{orcidlink}

\begin{document}

\title{Detector-Frame Quantum Limits for Heterodyne Timing in Relativistic Optical Links}

\author{Salman Sajad Wani\,\orcidlink{ 0000-0002-5262-9738}}
\affiliation{Qatar Center for Quantum Computing, College of Science and Engineering, Hamad Bin Khalifa University, Doha, Qatar}  

\author{Aatif Kaisar Khan\,\orcidlink{0000-0002-9305-3835}}
\email{aatifkaisarkhan@ufl.edu} 
\affiliation{Department of Physics, University of Florida, Gainesville, FL 32611-8440, USA}

\author{Saif Al-Kuwari}
\affiliation{Qatar Center for Quantum Computing, College of Science and Engineering, Hamad Bin Khalifa University, Doha, Qatar.}

\author{Mir Faizal}
\affiliation{Canadian Quantum Research Center, 204-3002 32 Ave Vernon, BC V1T 2L7 Canada}
\affiliation{Irving K. Barber School of Arts and Sciences, University of British Columbia - Okanagan, Kelowna, British Columbia V1V 1V7, Canada}
\affiliation{Department of Mathematical Sciences, Durham University, Upper Mountjoy, Stockton Road, Durham DH1 3LE, UK}
\affiliation{Faculty of Sciences, Hasselt University, Agoralaan Gebouw D, Diepenbeek, 3590 Belgium}

\begin{abstract}
Long-baseline optical links for space interferometry and satellite quantum networks rely on heterodyne phase tracking under Doppler motion and gravitational frequency shifts. These relativistic shifts can generate large detector-frame beat-frequency offsets over finite readout windows and thereby constrain the temporal precision of front-end phase readout. We develop a detector-frame framework for this problem using an effective two-mode signal--local-oscillator model in which relativistic motion rescales the local-oscillator frequency. Within this setting we derive closed-form Mandelstam--Tamm and Margolus--Levitin bounds for representative Gaussian resources and connect global state distinguishability with the operational precision of heterodyne phase tracking. In the bright-local-oscillator regime we identify a geometric--operational mismatch: the joint optical state can evolve on arbitrarily short global timescales while the relevant information rate is set by the measured detector record. Combining this result with an open-loop transient-acquisition analysis, we show that white phase noise and relativistic beat-frequency drift that is approximately constant over the acquisition window generate a cube-root scaling benchmark for the optimal pre-lock open-loop integration window and the minimum residual phase error. These results establish detector-frame benchmarks for quantum-limited heterodyne timing in relativistic optical links and clarify how global dynamical bounds enter task-restricted continuous optical measurements.
\end{abstract}

\maketitle

\section{Introduction}

Balanced heterodyne detection resolves relative optical phase by interfering a weak
signal with a bright local oscillator (LO) on a 50:50 beam splitter and monitoring
the difference photocurrent \cite{MandelWolf1995,Caves1981}. At optical carriers
($\omega \sim 10^{15}\,\mathrm{rad\,s}^{-1}$), such continuous-variable phase
tracking underlies precision timing and phase metrology in both tabletop
interferometers and large-scale space-based instruments
\cite{diels2006ultrashort,Schnabel2017,Andersen2016}. Space-based architectures such as
LISA employ heterodyne laser links and clock synchronization across moving spacecraft
with time-dependent MHz-scale beat notes set by orbital Doppler shifts and frequency planning
\cite{LISA2017,Tinto2021,TintoDhurandhar2014}, while satellite quantum links provide
a complementary long-baseline platform in which relativistic and gravitational
effects influence quantum optical signal propagation
\cite{Sidhu2021,Liao2017,Yin2020,Bruschi2014,Rideout2012}. In these systems the
target heterodyne offset frequency $\Delta\omega$ is typically much smaller than
the LO carrier frequency $\omega_{\mathrm{LO}}$ ($|\Delta\omega|\ll
\omega_{\mathrm{LO}}$). As a consequence, even modest relativistic frame shifts at
the carrier are amplified by the carrier-to-beat ratio into large absolute
local beat-note drifts. If uncompensated over the relevant acquisition window, such drift enters the front-end phase readout
prior to downstream digital correction. It therefore constrains frequency planning
and temporal phase-tracking performance, particularly during acquisition and related
transient regimes in which the phasemeter has not yet removed the offset through
active loop tracking
\cite{Tinto2021,TintoDhurandhar2014,Shaddock_2006,Bachman2017GRACEFO}. The focus of
the present work is the temporal precision of the intermediate-frequency (IF)
phase record used in front-end tracking and synchronization tasks rather than
generic optical ranging precision.

To analyze these timing limitations, it is natural to connect parameter-estimation
theory with quantum speed limits (QSLs)
\cite{Mandelstam1945,DeffnerCampbell2017,MargolusLevitin1998}. For unitary parameter
families generated by a time-independent Hamiltonian, the generator variance
controls the local statistical distinguishability relevant for parameter
estimation, while Hamiltonian moments also enter standard QSL bounds
\cite{BraunsteinCaves1994,Giovannetti2006,DeffnerCampbell2017}. These connections
establish a bridge between dynamical state evolution and metrological sensitivity,
although the operational precision ultimately depends on the measurement channel.
Recent work has developed relativistic QSL bounds for Gaussian systems, together
with prospective experimental tests, in Ref.~\cite{WaniKhanAlKuwariFaizal2025PLA},
while complementary treatments of relativistic quantum metrology and proper-time
interferometry appear in
Refs.~\cite{AhmadiBruschiFuentes2014SciRep,ZychCostaPikovskiBrukner2011NatCommun}.
Balanced heterodyne detection introduces an additional structural element:
although the signal and LO evolve independently under the free optical Hamiltonian,
physically accessible information arises only after interference and measurement,
where the scalar difference-photocurrent record encodes the relative phase. A central question in this setting is therefore to connect the detector-frame
state-distinguishability structure captured by QSL or quantum Fisher information
methods with the accumulated beat phase and finite-window temporal
information actually extracted by a balanced heterodyne receiver.  The present work develops this detector-frame bridge
as a quantum-technology benchmark for relativistic heterodyne timing in
long-baseline optical and satellite-link architectures. The analysis applies most
directly to effective heterodyne configurations in which the relativistically
shifted optical mode serves as the bright reference entering the interference
stage, while extensions to configurations in which the shifted mode is the weak
received field remain an important direction for future work.

The present work contributes three linked advances. First, it formulates
detector-frame quantum limits for balanced heterodyne readout in relativistic
optical links. Second, it identifies a bright-local-oscillator mismatch between
global state distinguishability and task-restricted detector performance. Third,
it derives an operational cube-root scaling law for open-loop transient phase
tracking under locally quasi-static relativistic beat-frequency drift, thereby providing a detector-frame benchmark
for timing and synchronization analysis in long-baseline photonic platforms under the stated finite-window validity conditions.

We construct a detector-frame QSL framework for balanced heterodyne readout in the
weak-field nonrelativistic regime ($v/c\equiv\beta\ll1$). Modeling the signal and
LO as two harmonic modes, we incorporate longitudinal Doppler and leading
Schwarzschild shifts through a detector-frame rescaling of the LO frequency. The
resulting description functions as an effective narrowband two-mode model: within
the weak-field adiabatic regime relevant to optical links, observer-dependent mode
mixing and particle-creation effects belong to the underlying curved-spacetime
field-theory background, while their contribution is assumed to be subleading on
the readout scales of interest
\cite{BirrellDavies1982,Unruh1976,Crispino2008,Bruschi2014}. For representative
Gaussian product inputs---coherent $\otimes$ coherent, squeezed-vacuum $\otimes$
coherent, and a reduced single-mode squeezed-coherent resource model---we derive
closed-form Mandelstam--Tamm and Margolus--Levitin bounds and use a fixed
Bures-angle threshold to define the corresponding distinguishability times. A
central conceptual result is the emergence of a geometric--operational mismatch in
the bright-LO limit: as the LO amplitude increases, global quantum speed limits
derived from joint-state distinguishability need not coincide with the operational
timing bounds relevant for heterodyne phase readout. In this regime the effective
information rate is therefore determined by the measurement channel rather than by
global optical-state separation alone.

At the operational level, the limiting factor shifts from global state-space motion
to finite-window estimation of the IF phase record. Within an untracked open-loop
transient-acquisition model prior to active phase locking, we express the
finite-window phase uncertainty in amplitude-spectral-density form and place it
within the framework of continuous-time estimation theory
\cite{BendatPiersol2010,Saulson2017,Tsang2011}. Balancing stochastic phase noise
against residual beat-frequency drift $\delta\omega$ yields an optimal
pre-processed integration window, for which the minimum residual phase error
obeys a cube-root drift scaling relation,
\(
\sqrt{\mathrm{MSE}_{\min}} \propto |\delta\omega|^{1/3}.
\)
The same detector-frame timing model then produces a squeezing--kinematic
break-even condition for these temporal IF-resolution tasks. Within the reduced
single-mode resource-scaling model considered here, rather than a full image-band-resolved phase-insensitive heterodyne noise model, and assuming a compressive
frequency plan, a representative low-Earth-orbit example yields an effective
requirement of about $10.6$~dB of detected continuous-variable squeezing together
with an absolute best-case transmissivity floor of $\eta\ge0.913$. Comparison with
reported atmospheric link losses places this benchmark well outside the regime of
conventional ground-to-space optical channels \cite{Liao2017}. The worked temporal
example considered later should therefore be interpreted as an illustrative
detector-front-end timing benchmark for local relativistic optical-link operating points, rather than as
a complete orbital, tracking, or engineering model of a specific mission phasemeter.

\section{Quantum Speed Limit for Two-Mode Product States}
\label{sec:QSL-two-mode}

To formulate the global detector-frame state-geometry layer of the problem, we model
the signal ($s$) and local oscillator (LO) as a bipartite harmonic system governed by
the free Hamiltonian \cite{MandelWolf1995,Leonhardt1997,WallsMilburn2008}
\begin{equation}
\hat{H}
=
\hbar \omega_s \left(\hat{a}_s^{\dagger}\hat{a}_s + \tfrac{1}{2}\right)
+
\hbar \omega_{\mathrm{LO}} \left(\hat{a}_{\mathrm{LO}}^{\dagger}\hat{a}_{\mathrm{LO}} + \tfrac{1}{2}\right).
\label{eq:Hfree}
\end{equation}
This section concerns the global distinguishability structure generated by the
effective two-mode Hamiltonian. The task-restricted detector readout enters later in
Secs.~\ref{sec:time-resolution} and \ref{sec:applications_main}.

Within this framework, we evaluate three Gaussian product families: coherent
$\otimes$ coherent, squeezed-vacuum $\otimes$ coherent, and a reduced single-mode
squeezed-coherent resource model $\otimes$ coherent
\cite{Glauber1963,Walls1983,Yuen1976}. The first two families provide exact
reference cases for the detector-frame QSL geometry, while the third will later serve
as a reduced resource-scaling model for the temporal readout discussion.

In the heterodyne setting, the signal and the local oscillator are coupled operationally at the detector through interference on the beam splitter, and the measured photocurrent depends on both modes. In the present section we first isolate the global state-geometry layer, which is generated by the free two-mode Hamiltonian $\hat H=\hat H_s+\hat H_{\mathrm{LO}}$. For product inputs this evolution factorizes prior to measurement, while the later heterodyne readout mixes the modes through the detection observable. The resulting global quantum speed limit is therefore set by the energy statistics of the full two-mode system. For a bright local oscillator, the LO photon number is large, so its contribution can dominate the total energy variance $(\Delta H)^2$ and strongly influence the Mandelstam--Tamm timescale. A detector-frame shift $\omega_{\mathrm{LO}}\to\omega'_{\mathrm{LO}}=\kappa\omega_{\mathrm{LO}}$ then rescales the LO Hamiltonian contribution and modifies the corresponding global QSL \cite{DeffnerCampbell2017}.

For a pure bipartite initial state $\lvert \psi_0\rangle$ evolving dynamically to
$\lvert \psi_T\rangle$ over an interrogation time $T$, state distinguishability is
quantified by two geometric metrics: the fidelity
$F(T)=\lvert \langle \psi_0 \mid \psi_T\rangle\rvert^2$ and the corresponding Bures
angle $\mathcal L(T)=\arccos\!\sqrt{F(T)}$
\cite{Bures1969,Uhlmann1976,Jozsa1994}. Extracting the zero-point vacuum energy
$
E_0=\frac{\hbar}{2}\,(\omega_s+\omega_{\mathrm{LO}})
$
together with the initial energy moments $\langle H\rangle$ and $\Delta H$, which for
the time-independent Hamiltonian \eqref{eq:Hfree} remain constant along the unitary
evolution generated by $\hat H$, the generalized Mandelstam-Tamm (MT) and
Margolus-Levitin (ML) timescales bound the dynamics via
\cite{Mandelstam1945,MargolusLevitin1998,DeffnerCampbell2017}
\begin{equation}
\mathcal T_{\mathrm{MT}}(T)=\frac{\hbar\,\mathcal L(T)}{\Delta H},
\qquad
\mathcal T_{\mathrm{ML}}(T)=\frac{\hbar\,[1-\cos\mathcal L(T)]}{\langle H\rangle-E_0},
\label{eq:TMT_TML_main}
\end{equation}
yielding the ultimate global QSL time
$
\tau_{\mathrm{QSL}}(T)=\max\!\left\{\mathcal T_{\mathrm{MT}}(T),\mathcal T_{\mathrm{ML}}(T)\right\}.
$
For the product-state families considered here, and in the bright-LO regime at fixed evolution horizon, the MT contribution to this bound decreases asymptotically as the inverse local-oscillator amplitude, while the ML contribution decreases faster.

\textit{Coherent$\otimes$coherent.}-For a bipartite coherent preparation $\lvert \psi_0\rangle=\lvert \alpha_s\rangle\otimes\lvert \alpha_{\mathrm{LO}}\rangle$, the fidelity strictly factorizes into independent Gaussian overlaps \cite{Glauber1963,Perelomov1986,MandelWolf1995}:
\begin{equation}
F_{\text{cc}}(T)=
\exp\!\Bigl[
-4|\alpha_s|^2\sin^2\!\bigl(\tfrac{\omega_s T}{2}\bigr)
-4|\alpha_{\mathrm{LO}}|^2\sin^2\!\bigl(\tfrac{\omega_{\mathrm{LO}}T}{2}\bigr)
\Bigr],
\end{equation}
with the composite energy moments evaluating to
\begin{subequations}
\begin{align}
\langle H\rangle_{\text{cc}} &=
\hbar\omega_s\Bigl(|\alpha_s|^2+\tfrac{1}{2}\Bigr)
+\hbar\omega_{\mathrm{LO}}\Bigl(|\alpha_{\mathrm{LO}}|^2+\tfrac{1}{2}\Bigr), \\[1em]
(\Delta H)^2_{\text{cc}} &=\hbar^2\bigl(\omega_s^2|\alpha_s|^2+\omega_{\mathrm{LO}}^2|\alpha_{\mathrm{LO}}|^2\bigr).
\end{align}
\end{subequations}

\textit{Squeezed-vacuum$\otimes$coherent.}-Alternatively, injecting a squeezed vacuum into the signal port, $\lvert \psi_0\rangle=\hat S_s(r,0)\lvert 0\rangle\otimes\lvert \alpha_{\mathrm{LO}}\rangle$, yields the modified fidelity \cite{Walls1983,Schumaker1986,BarnettRadmore2002}
\begin{equation}
F_{\text{sv}\otimes\text{c}}(T)=
\frac{
\exp\!\Bigl[-4|\alpha_{\mathrm{LO}}|^2\sin^2\!\bigl(\tfrac{\omega_{\mathrm{LO}}T}{2}\bigr)\Bigr]
}{
\sqrt{1+\sinh^2(2r)\,\sin^2(\omega_s T)}
}.
\end{equation}
The corresponding energy moments reflect the non-classical variance of the squeezed mode:
\begin{subequations}
\begin{align}
\langle H\rangle_{\text{sv}\otimes\text{c}} &=
\hbar\omega_s\Bigl(\sinh^2 r+\tfrac{1}{2}\Bigr)
+\hbar\omega_{\mathrm{LO}}\Bigl(|\alpha_{\mathrm{LO}}|^2+\tfrac{1}{2}\Bigr), \\[1em]
(\Delta H)^2_{\text{sv}\otimes\text{c}}
&=\hbar^2\bigl(2\omega_s^2\sinh^2 r\,\cosh^2 r+\omega_{\mathrm{LO}}^2|\alpha_{\mathrm{LO}}|^2\bigr).
\end{align}
\end{subequations}

\textit{Reduced single-mode squeezed-coherent resource model$\otimes$coherent.}-For
the fully displaced signal
$\lvert \psi_0\rangle=\hat D_s(\alpha_s)\hat S_s(r,0)\lvert 0\rangle\otimes\lvert \alpha_{\mathrm{LO}}\rangle$,
the fidelity $F_{\text{sc}\otimes\text{c}}(T)$ is strictly analytic
\cite{Yuen1976,Scutaru1998,Banchi2015}. In the later operational discussion, this
family is used as a reduced single-mode resource-scaling model for how quadrature
suppression enters the temporal readout estimate.  Restricting the configuration such that the coherent displacement $\alpha_s$ aligns exactly with the squeezed quadrature, the energy moments evaluate to:
\begin{subequations}
\begin{align}
\langle H\rangle_{\text{sc}\otimes\text{c}} &=
\hbar\omega_s\Bigl(|\alpha_s|^2+\sinh^2 r+\tfrac{1}{2}\Bigr)
+\hbar\omega_{\mathrm{LO}}\Bigl(|\alpha_{\mathrm{LO}}|^2+\tfrac{1}{2}\Bigr), \\[1em]
(\Delta H)^2_{\text{sc}\otimes\text{c}}&=\hbar^2\Bigl[
\omega_s^2\bigl(2\sinh^2 r\,\cosh^2 r+|\alpha_s|^2 e^{-2r}\bigr)
+\omega_{\mathrm{LO}}^2|\alpha_{\mathrm{LO}}|^2
\Bigr].
\end{align}
\end{subequations}

Appendix~\ref{app:two-mode-QSL} details the intermediate algebraic derivations and provides the explicit analytic form of the fidelity for the reduced squeezed-coherent model used in this section.

\section{Relativistic Frequency Shifts and Quantum Speed Limits}
\label{sec:relativistic}

To embed the global two-mode QSL problem in the detector frame, we next specify how
relativistic kinematic shifts enter the effective signal-LO Hamiltonian. All
timescales are evaluated with respect to the detector proper time governing the
measured photocurrent record. We take the signal field and detector to define the
local operational frame, so that $\omega_s$ is specified there, while the remote LO
acquires detector-frame corrections from relative motion and gravitational potential
offset. Within the weak-field, narrowband, adiabatic regime considered here, the
dominant observer dependence is modeled by a scalar rescaling of the LO frequency.
This effective detector-frame construction feeds the global distinguishability
analysis of the present section; the task-restricted temporal readout consequences are
developed later in Secs.~\ref{sec:time-resolution} and
\ref{sec:applications_main}.

For a strictly longitudinal relative velocity $v_\parallel \equiv c\beta_\parallel$ established between the remote LO source and the local detector, expanding the special-relativistic Doppler factor to leading order in the non-relativistic limit ($|\beta_\parallel|\ll 1$) yields the exact kinematically rescaled frequency \cite{Ashby2003LRR}:
\begin{equation}
\omega_{\mathrm{LO}}^{(\mathrm{D})}
=
\omega_{\mathrm{LO}}
\sqrt{\frac{1+\beta_\parallel}{1-\beta_\parallel}}
\simeq
\omega_{\mathrm{LO}}(1+\beta_\parallel).
\label{eq:Doppler_longitudinal}
\end{equation}

Parallel to this kinematic boost, the gravitational frequency shift is taken to be the weak-field potential difference between the emission and reception points \cite{Ashby2003LRR}. Assuming the detector is on the Earth's surface at radius $R_{\oplus}$ and the source is at orbital radius $r_{\mathrm{orb}}$, a descending downlink falls into the potential well and undergoes a gravitational blueshift. This shift enters at leading order as:
\begin{equation}
\omega_{\mathrm{LO}}^{(\mathrm{G})}
\simeq
\omega_{\mathrm{LO}}\!\left[1+\frac{GM}{c^2}\left(\frac{1}{R_{\oplus}}-\frac{1}{r_{\mathrm{orb}}}\right)\right].
\label{eq:grav_weakfield}
\end{equation}

Combining these precise first-order corrections yields the combined macroscopic frame-rescaling factor $\omega_{\mathrm{LO}}'=\kappa \omega_{\mathrm{LO}}$, with the parameter defined rigorously as
$
\kappa \simeq 1+\beta_\parallel+\frac{\Delta\Phi}{c^2},
$
where $\Delta\Phi$ represents the positive magnitude of the relevant gravitational potential difference (e.g., $\Delta\Phi = GM(1/R_{\oplus}-1/r_{\mathrm{orb}})$ for a space-to-ground downlink, or the solar potential difference for a heliocentric deep-space link).

Modeling the frame transformation through the scalar replacement $(\omega_{\mathrm{LO}}\mapsto \kappa\omega_{\mathrm{LO}})$ amounts to adopting an effective single-mode description of the LO in the detector frame. In relativistic quantum field theory \cite{Bruschi2014,BirrellDavies1982}, changes of observer or background can in general induce mode mixing through a Bogoliubov transformation,
$
\hat a_{\Omega}'=\int d\omega\left(\alpha_{\Omega\omega}\hat a_\omega+\beta_{\Omega\omega}^{*}\hat a_\omega^{\dagger}\right),
$
so the scalar treatment used here should be read as an approximation valid when a narrowband LO wavepacket remains sharply concentrated in one effective detector-frame mode and the off-diagonal mixing is negligible on the scales relevant to the readout model.

In the regime relevant to the present optical links, acceleration-induced vacuum excitation is enormously suppressed. Operator mixing and spontaneous particle creation from the vacuum are controlled by the off-diagonal coefficients $|\beta_{\Omega\omega}|^2$; at optical carrier frequencies ($\omega_{\mathrm{LO}}\sim 10^{15}~\mathrm{rad/s}$), the associated Unruh-type thermal factor scales as $\exp(-2\pi c\,\omega_{\mathrm{LO}}/a)$, where $a$ denotes the local proper acceleration \cite{Unruh1976,Crispino2008}. For orbiting satellites, the free-fall setting removes this particular acceleration scale from the link model, and for stationary terrestrial accelerations ($a\approx 9.8~\mathrm{m/s^2}$) the exponent is astronomically large in magnitude, so acceleration-induced vacuum admixture is negligible for the optical frequencies considered here.

The scalar frequency-shift model therefore rests on a restricted adiabatic regime defined by three ingredients: non-relativistic relative velocities ($|\beta_\parallel|\ll 1$), weak-field backgrounds ($\Phi/c^2\ll 1$), and optical carriers for which acceleration-induced vacuum excitation is negligible on the timescales of interest. Within this regime, we approximate the LO by a single effective detector-frame mode whose dominant transformation consists of a frequency rescaling together with an overall phase, schematically $(\hat a_{\Omega}'\approx e^{i\Theta}\hat a_{\kappa\omega})$ \cite{BirrellDavies1982,Bruschi2014}. This effective description is sufficient for the two-mode Hamiltonian model adopted below.

Under this effective single-mode approximation for the LO, the two-mode generator expressed in detector time takes the form
\begin{equation}
\hat{H}'
=
\hbar\omega_s\Bigl(\hat{a}_s^\dagger \hat{a}_s+\tfrac{1}{2}\Bigr)
+\hbar\omega_{\mathrm{LO}}'\Bigl(\hat{a}_{\mathrm{LO}}^\dagger \hat{a}_{\mathrm{LO}}+\tfrac{1}{2}\Bigr).
\label{eq:Hprime}
\end{equation}

Relative to the unshifted baseline, the local-oscillator contribution to the mean energy scales linearly with $\kappa$, while its contribution to the energy variance scales quadratically. Consequently, within the effective detector-frame model, the generalized Mandelstam-Tamm and Margolus-Levitin expressions for the bipartite product inputs of Sec.~\ref{sec:QSL-two-mode} retain the same analytical form after the substitution $\omega_{\mathrm{LO}}\mapsto \omega_{\mathrm{LO}}'$.

For the coherent$\otimes$coherent and squeezed-vacuum$\otimes$coherent families,
together with the reduced single-mode squeezed-coherent resource model introduced in
Sec.~\ref{sec:QSL-two-mode}, the global fidelity $F(T)$ retains the product structure
of independent mode evolutions, while the local-oscillator overlap factors absorb the
scaled kinematic replacement $\omega_{\mathrm{LO}}T\mapsto\omega_{\mathrm{LO}}'T$
without requiring additional operator mixing.

The global QSL time $\tau_{\mathrm{QSL}}(T)$ inherits observer dependence through two channels: the rescaled energy statistics entering the Hamiltonian moments, and the modified Bures angle $\mathcal L(T)=\arccos\!\sqrt{F(T)}$ through the shifted LO overlap. Within the present effective model, an approaching longitudinal blue shift ($\kappa>1$) increases the LO energy scale in detector time and typically reduces the corresponding global QSL timescale for fixed state parameters, while a redshifted configuration ($\kappa<1$) produces the opposite trend. The operational consequences for heterodyne phase and timing resolution are developed separately in Secs.~\ref{sec:time-resolution} and \ref{sec:applications_main}, where the task-restricted measurement model is introduced explicitly. A possible extension to asymmetric links would assign distinct kinematic factors $(\kappa_s,\kappa_{\mathrm{LO}})$ to the two modes.


\section{Quantum Time Resolution and Task-Restricted Limits in Heterodyne Detection}
\label{sec:time-resolution}

This section provides the bridge between the global detector-frame QSL geometry of
Secs.~\ref{sec:QSL-two-mode} and \ref{sec:relativistic} and the task-restricted
temporal precision of the heterodyne readout. Fixing a Bures-angle threshold defines
a QSL hitting time in the detector-frame state geometry
\cite{Bures1969,Uhlmann1976,Jozsa1994,DeffnerCampbell2017}. Using
the detector-frame accumulated phase map introduced below, we map that hitting time
to an associated detector-frame beat-phase scale. We let $\mathcal L(t)$ denote the
detector-frame Bures angle between $\lvert\psi_0\rangle$ and $\lvert\psi_t\rangle$
under $\hat H'$. For a chosen $\mathcal L_\star\in(0,\pi/2]$ (where
$F_\star=\cos^2\mathcal L_\star$), we define the first hitting time by
\begin{equation}
t_\star \equiv \inf\{t\ge 0:\ \mathcal L(t)\ge \mathcal L_\star\}.
\label{eq:first_hitting_time}
\end{equation}

The Mandelstam--Tamm and Margolus--Levitin bounds yield a finite-angle unitary QSL
written in terms of the same Bures-angle threshold
\cite{Mandelstam1945,MargolusLevitin1998,GiovannettiLloydMaccone2003,DeffnerCampbell2017},
\begin{equation}
t_\star \ge
\max\!\left\{
\frac{\hbar\,\mathcal L_\star}{\Delta H'},
\frac{\hbar\,(1-\cos\mathcal L_\star)}{\langle H'\rangle-E_0'}
\right\},
\label{eq:tstar_QSL_bound}
\end{equation}
where \(E_0'\) is the ground energy of the local frozen two-mode Hamiltonian
specified below.The second term in Eq.~\eqref{eq:tstar_QSL_bound} is the
finite-angle ML-type form for a prescribed fidelity or Bures-angle threshold,
used here as a lower bound on the two-oscillator hitting time.
For the present model, Eq.~\eqref{eq:tstar_QSL_bound} is applied to the local frozen
detector-frame Hamiltonian. Thus the detector-frame frequencies are evaluated at the
chosen epoch, and their slow macroscopic drift is not included inside the short-time
generator used in Eq.~\eqref{eq:tstar_QSL_bound}. The two-mode oscillator
Hamiltonian is

\begin{equation}
\hat H'
=
\hbar\omega_s\left(\hat n_s+\frac{1}{2}\right)
+
\hbar\omega_{\mathrm{LO}}'
\left(\hat n_{\mathrm{LO}}+\frac{1}{2}\right),
\label{eq:two_mode_H_specialized}
\end{equation}

acting on the tensor-product Hilbert space
$\mathcal H_s\otimes\mathcal H_{\mathrm{LO}}$. Hence its ground energy is
\begin{equation}
E_0'
=
\frac{\hbar}{2}
\left(\omega_s+\omega_{\mathrm{LO}}'\right),
\label{eq:two_mode_ground_energy}
\end{equation}
and the energy above the ground state entering the ML denominator is
{
\begin{equation}
\langle H'\rangle-E_0'
=
\hbar\omega_s\langle \hat n_s\rangle
+
\hbar\omega_{\mathrm{LO}}'
\langle \hat n_{\mathrm{LO}}\rangle .
\label{eq:two_mode_energy_above_ground}
\end{equation}
}
{Similarly, the energy variance entering the MT denominator is fixed by the
additive structure of the two-mode Hamiltonian:}
{
\begin{equation}
(\Delta H')^2
=
(\Delta H_s)^2
+
(\Delta H_{\mathrm{LO}}')^2
+
2\,{\rm Cov}(\hat H_s,\hat H_{\mathrm{LO}}'),
\label{eq:two_mode_variance_general}
\end{equation}
}
{where}
{
\begin{equation}
{\rm Cov}(\hat H_s,\hat H_{\mathrm{LO}}')
=
\frac{1}{2}
\langle
\hat H_s\hat H_{\mathrm{LO}}'
+
\hat H_{\mathrm{LO}}'\hat H_s
\rangle
-
\langle \hat H_s\rangle
\langle \hat H_{\mathrm{LO}}'\rangle .
\label{eq:two_mode_covariance}
\end{equation}
}
{For the product signal--LO states used in the main estimates, this covariance
vanishes, giving}
{
\begin{equation}
(\Delta H')^2
=
(\Delta H_s)^2
+
(\Delta H_{\mathrm{LO}}')^2 .
\label{eq:two_mode_variance_product}
\end{equation}
}
{For coherent product states, these expressions reduce to}
{
\begin{equation}
\langle H'\rangle-E_0'
=
\hbar\omega_s|\alpha_s|^2
+
\hbar\omega_{\mathrm{LO}}'|\alpha_{\mathrm{LO}}|^2 ,
\label{eq:coherent_energy_above_ground}
\end{equation}
}
{and}
{
\begin{equation}
(\Delta H')^2
=
\hbar^2\omega_s^2|\alpha_s|^2
+
\hbar^2(\omega_{\mathrm{LO}}')^2
|\alpha_{\mathrm{LO}}|^2 .
\label{eq:coherent_energy_variance}
\end{equation}
}
{Thus Eq.~\eqref{eq:tstar_QSL_bound} is applied to explicitly specified energy
moments of the additive two-mode oscillator Hamiltonian
(App.~\ref{app:two-mode-QSL}).}

{Because this oscillator model is exactly solvable, we also distinguish the general
MT/ML lower bound from the exact detector-frame distinguishability.} {For product
initial states,}
{
\begin{equation}
F(t)
=
\left|
\langle\psi_0|
e^{-i\hat H't/\hbar}
|\psi_0\rangle
\right|^2
=
F_s(t)F_{\mathrm{LO}}(t).
\label{eq:two_mode_fidelity_factorization}
\end{equation}
}
{For coherent product states, the exact fidelity is}
{
\begin{equation}
F(t)
=
\exp\!\left[
-4|\alpha_s|^2
\sin^2\!\left(\frac{\omega_s t}{2}\right)
-
4|\alpha_{\mathrm{LO}}|^2
\sin^2\!\left(\frac{\omega_{\mathrm{LO}}' t}{2}\right)
\right],
\label{eq:two_mode_coherent_fidelity}
\end{equation}
}
{and the actual Bures angle is}
{
\begin{equation}
\mathcal L(t)=\arccos\sqrt{F(t)}.
\label{eq:bures_from_exact_fidelity}
\end{equation}
}
{Thus Eq.~\eqref{eq:tstar_QSL_bound} gives a lower bound on the hitting time,
whereas the exact factorized fidelity determines the model-specific Bures angle
used to verify the scaling of \(t_\star\) (App.~\ref{app:two-mode-QSL}).}

{Phase-rate formulations of QSLs provide the state-level context for this construction:
they relate QSLs to the rate of change of a quantum-state phase or state overlap,
with the MT branch governed by the projective-state speed $\Delta H/\hbar$
\cite{Sun2021PhaseQSL}.} {In the present signal--LO model, the combined two-mode state
evolves under $\hat H'$, so the QSL hitting time $t_\star$ is a state-space
distinguishability time of the joint optical state.}

{The heterodyne phase used below has the same Hamiltonian origin but is a different
phase functional.} {Balanced detection measures the detector-frame relative
coherent-amplitude phase
\(\psi(t)=\psi_s(t)-\psi_{\rm LO}'(t)\) through the interference operator
\(\hat a_s^\dagger\hat a_{\rm LO}+\hat a_{\rm LO}^\dagger\hat a_s\).} {This measured
phase is distinct from the full state-overlap phase
\(\arg\langle\psi_0|\psi_t\rangle\).} {For free local
evolution, the measured beat-phase rate is $|d\psi/dt|=|\Delta\omega'|$, up to the
chosen sign convention.} {We therefore use the state-level QSL to define $t_\star$,
and then map that interval to the beat phase accumulated by the heterodyne readout.}

{For a detector-frame heterodyne measurement, the observable beat phase is the
accumulated difference between the signal phase and the detector-frame
local-oscillator phase.} {In a realistic optical link the beat frequency need not be
globally constant, because the frame factor, residual Doppler shift, gravitational
contribution, and tracking offsets can vary during the observation.} {We therefore
write the beat phase from a local epoch $t_0$ as}
{
\begin{equation}
\psi(t_0+T)
=
\psi(t_0)
+
\int_0^T
\Delta\omega'(t_0+u)\,\mathrm{d}u ,
\qquad
\Delta\omega'(t)
=
\omega_s(t)-\omega_{\rm LO}'(t).
\label{eq:psi_integral_general}
\end{equation}
}

{The phase in Eq.~\eqref{eq:psi_integral_general} is a relative detector-frame
phase.} {A common phase convention applied to both the received signal and the local
oscillator leaves $\psi(t)=\psi_s(t)-\psi_{\rm LO}'(t)$ unchanged.} {Rotating-frame
or demodulation conventions can move the same relative phase between amplitudes,
Hamiltonian terms, tracking references, and readout operators, but the consistently
transformed photocurrent contains the same measured beat phase.}

{The linear beat-phase model is only the locally stationary limit of
Eq.~\eqref{eq:psi_integral_general}.} {If
$\Delta\omega'(t_0+u)\simeq\Delta\omega'_0$, with
$\Delta\omega'_0\equiv\Delta\omega'(t_0)$, over the relevant estimation window, then}
{
\begin{equation}
\psi(t_0+T)
\simeq
\psi(t_0)+\Delta\omega'_0 T .
\label{eq:psi_linear_local}
\end{equation}
}
{The local approximation is valid only when the chirp contribution
$\frac12\dot{\Delta\omega}'_0T^2$ is small compared with the phase scale being
resolved.} {The QSL-associated detector beat-phase scale is therefore the accumulated beat phase
during the QSL hitting interval,}
{
\begin{equation}
\Psi_\star(t_0)
\equiv
\left|
\int_0^{t_\star}
\Delta\omega'(t_0+u)\,\mathrm{d}u
\right|.
\label{eq:Psi_star_integral}
\end{equation}
}
{In the locally stationary limit this reduces to}
{
\begin{equation}
\Psi_\star(t_0)
\simeq
|\Delta\omega'_0|t_\star ,
\qquad
\Delta\omega'_0\equiv\Delta\omega'(t_0),
\label{eq:Psi_star_local}
\end{equation}
}
{and Eq.~\eqref{eq:tstar_QSL_bound} gives the local estimate}
{
\begin{equation}
\Psi_\star(t_0)
\gtrsim
|\Delta\omega'_0|
\max\!\left\{
\frac{\hbar\,\mathcal L_\star}{\Delta H'},
\frac{\hbar\,(1-\cos\mathcal L_\star)}{\langle H'\rangle-E_0'}
\right\},
\label{eq:psi_star_bound}
\end{equation}
}
{up to corrections of order $\dot{\Delta\omega}'_0t_\star^2$.}

In practical phasemeter architectures, heterodyne-frequency planning and front-end
readout constraints are engineered to keep the beat note within the usable
measurement band \cite{Heinzel2004,Barke_2015,Guzman_2011}. For the three Gaussian
product families considered in Sec.~\ref{sec:QSL-two-mode}, the LO contribution
enters the fidelity through the common factor
\begin{equation}
\exp\!\Bigl[-4|\alpha_{\mathrm{LO}}|^2
\sin^2\!\bigl(\tfrac{\omega_{\mathrm{LO}}' t}{2}\bigr)\Bigr].
\label{eq:common_LO_factor}
\end{equation}
At short times,
\begin{equation}
\exp\!\Bigl[-4|\alpha_{\mathrm{LO}}|^2
\sin^2\!\bigl(\tfrac{\omega_{\mathrm{LO}}' t}{2}\bigr)\Bigr]
=
\exp\!\Bigl[-|\alpha_{\mathrm{LO}}|^2(\omega_{\mathrm{LO}}' t)^2+O(t^4)\Bigr],
\label{eq:LO_short_time_expansion}
\end{equation}
while the corresponding signal-dependent prefactor is $1+O(t^2)$ with coefficients
independent of $|\alpha_{\mathrm{LO}}|$. Hence, for any fixed threshold
$F_\star\in(0,1)$ and fixed detector-frame frequencies, the first hitting time
satisfies
\begin{equation}
t_\star = O(|\alpha_{\mathrm{LO}}|^{-1})
\qquad
(|\alpha_{\mathrm{LO}}|\to\infty).
\label{eq:tstar_brightLO_scaling}
\end{equation}
{This asymptotic statement is made at fixed threshold, fixed signal resource, and
fixed locally frozen detector-frame frequencies over the QSL interval.}

{If the receiver frequency plan keeps the instantaneous detector-frame beat note
within a finite measurement band, $|\Delta\omega'(t)|\le\Omega_{\rm band}$, then the
exact accumulated QSL-associated phase obeys}
{
\begin{equation}
\Psi_\star(t_0)
\le
\int_0^{t_\star}
|\Delta\omega'(t_0+u)|\,\mathrm{d}u
\le
\Omega_{\rm band}t_\star .
\label{eq:Psi_star_band_bound}
\end{equation}
}
{Because Eq.~\eqref{eq:tstar_brightLO_scaling} gives
$t_\star=O(|\alpha_{\rm LO}|^{-1})$, it follows that}
{
\begin{equation}
\Psi_\star(t_0)\to0
\qquad
(|\alpha_{\rm LO}|\to\infty).
\end{equation}
}
{Thus the vanishing of the QSL-associated beat-phase scale does not require a
globally constant beat note; it follows from bounded detector-frame beat frequency
together with the bright-reference collapse of the global state-distinguishability
time.} The geometric state-separation timescale obtained from the joint optical state
is controlled by carrier-scale evolution ($\omega_{\mathrm{LO}}'$), whereas the
operational estimator is governed by the beat note ($\Delta\omega'$) after
interference and demodulation. {In the bright-reference limit, the global
distinguishability threshold becomes asymptotically separated from the operational
phase-tracking problem.} {The latter is therefore governed by task-restricted bounds
derived from the $\hat I_{\mathrm{diff}}$ record.}

{The phase parameter itself is classical after demodulation, but the phase sensitivity
is quantum-limited when the noise floor is set by the optical measurement statistics.}
{At the level of optimal quantum parameter estimation, the relative phase is represented
by a phase-encoded optical state family.} {The corresponding finite-window estimator
obeys the quantum Cram\'er--Rao bound
$
\Delta\psi(T)
\ge
\frac{1}{\sqrt{\mathcal F_Q^{(T)}(\psi)}} ,
$
where $\mathcal F_Q^{(T)}(\psi)$ is the total phase Fisher information available
in the ideal quantum-limited record over the measurement window $T$
\cite{Helstrom1976,BraunsteinCaves1994}.} {Under the local phase-time map
$d\psi/dt=\Delta\omega'(t_0)$, this gives}
{
\begin{equation}
    \mathcal F_Q^{(T)}(t)
=
|\Delta\omega'(t_0)|^2\mathcal F_Q^{(T)}(\psi),
\qquad
\Delta t(T)
\ge
\frac{1}
{|\Delta\omega'(t_0)|\sqrt{\mathcal F_Q^{(T)}(\psi)}} .
\end{equation}
}
{For a closed unitary time-translation family, this Fisher-information timing bound
reduces to the local Mandelstam--Tamm structure.} {In particular, for pure-state
evolution generated by the frozen detector-frame Hamiltonian $H'$,}
{
\begin{equation}
    \mathcal F_Q(t)
=
\frac{4(\Delta H')^2}{\hbar^2},
\qquad
\Delta t
\ge
\frac{\hbar}{2\Delta H'} .
\end{equation}
}
{Thus the phase-to-time QCRB is the task-restricted timing counterpart of the same
generator-variance structure that appears in the MT part of
Eq.~\eqref{eq:tstar_QSL_bound}.} {The ML part of Eq.~\eqref{eq:tstar_QSL_bound}
remains the complementary finite-angle closed-system constraint governed by the mean
excitation energy $\langle H'\rangle-E_0'$.}

{In the balanced heterodyne model, the relative phase is read out through
$
\hat I_{\rm diff}
=
\hat a_s^\dagger\hat a_{\rm LO}
+
\hat a_{\rm LO}^\dagger\hat a_s .
$}
{At a linearized operating point, local error propagation gives}
{
\begin{equation}
\Delta\psi_{\rm q}
=
\frac{\Delta I_{\rm diff}}
{\left|
\partial_{\delta\psi}
\langle \hat I_{\rm diff}\rangle
\right|},
\label{eq:phase_sensitivity_bridge_main}
\end{equation}
}
{where
$
\Delta I_{\rm diff}^2
=
\langle\hat I_{\rm diff}^2\rangle
-
\langle\hat I_{\rm diff}\rangle^2
$
is the quantum variance in the chosen input state.} {Thus $\Delta\psi_{\rm q}$ is
the quantum-noise-limited contribution to the phase sensitivity of the specified
heterodyne measurement.} {The balanced-detection expression in
Eq.~\eqref{eq:phase_sensitivity_bridge_main} is therefore the measurement-specific
sensitivity of the heterodyne readout.} {It coincides with the corresponding Fisher-information limit only when this
specified readout saturates that limit
(Apps.~\ref{app:quantum-time-res}, \ref{app:coh-coh-BHD}, and
\ref{app:sqcoh-BHD}).}

{For a continuous readout, the quantum difference-current noise is referred to the
input phase through the local slope
$
G_\psi
\equiv
\partial_{\delta\psi}\langle\hat I_{\rm diff}\rangle .
$}
{In the same normalization,}
{
\begin{equation}
S_{\psi,{\rm q}}
=
\frac{S_{I,{\rm q}}}{|G_\psi|^2}.
\label{eq:quantum_current_to_phase_psd}
\end{equation}
}
{Additional electronic, tracking, and technical noise contributions enter at the PSD
level, so the total demodulated phase-noise PSD may be written as}
{
\begin{equation}
S_\psi
=
S_{\psi,{\rm q}}
+
S_{\psi,{\rm tech}} .
\label{eq:total_phase_psd}
\end{equation}
}

{Here $\Psi_\star(t_0)$ is the QSL-associated beat-phase scale, whereas
$\Delta\psi(T)$ is the finite-window uncertainty of the demodulated phase estimate.}
{Continuous-wave interferometry is commonly reported through the one-sided phase-noise
ASD $S_\psi^{1/2}$, in units $\mathrm{rad}/\sqrt{\mathrm{Hz}}$.} {In the present
notation, $S_\psi$ denotes the total locally stationary phase-noise PSD of the
demodulated readout.} {It may be quantum-limited, $S_\psi=S_{\psi,{\rm q}}$, or may
contain additional technical readout contributions as in Eq.~\eqref{eq:total_phase_psd}.}
{If this PSD is approximately white over the averaging band, a rectangular average
over a window $T$ gives (App.\ref{F})}
{
\begin{align}
    \mathrm{Var}[\hat\psi_T]
=
\frac{S_\psi}{2T},
\qquad
\Delta\psi(T)
=
\frac{S_\psi^{1/2}}{\sqrt{2T}} .
\end{align}
}
{Thus $\Delta\psi(T)$ is the finite-time readout representation of the
phase-estimation uncertainty; it is quantum-limited only when the technical
contribution to $S_\psi$ is negligible.}

{The same normalization defines an equivalent ASD associated with a finite-window
uncertainty,}
{
\begin{equation}
S_{\psi,{\rm eq}}^{1/2}(T)
\equiv
\sqrt{2T}\,\Delta\psi(T),
\qquad
S_{t,{\rm eq}}^{1/2}(T)
\equiv
\sqrt{2T}\,\Delta t(T).
\label{eq:equiv_ASD_map_refined}
\end{equation}
}
{The corresponding local timing uncertainty is obtained from the same phase-time map,}
{
\begin{equation}
\Delta t(T)
=
\frac{\Delta\psi(T)}
{\left|d\psi/dt\right|_{t_0}}
=
\frac{\Delta\psi(T)}
{|\Delta\omega'(t_0)|},
\label{eq:local_time_uncertainty_from_phase}
\end{equation}
}
{provided that the beat frequency is locally stationary over the averaging window.} {If
$\Delta\psi(T)$ is set by the quantum-limited phase sensitivity, this gives}
{
\begin{equation}
    \Delta t_{\rm q}(T)
=
\frac{1}{|\Delta\omega'(t_0)|}
\sqrt{\frac{S_{\psi,{\rm q}}}{2T}}.
\end{equation}
}
{If technical noise is present, $S_{\psi,{\rm q}}$ is replaced by the total $S_\psi$.}
{These results supply the quantities used in the operational benchmark analysis
of Sec.~\ref{sec:applications_main}: the QSL construction fixes the
state-distinguishability scale, while the finite-window phase and timing
uncertainties characterize the task-restricted heterodyne readout.}

\section{Operational Benchmarks and Hardware Constraints}
\label{sec:applications_main}

The Gaussian overlap formulas and associated asymptotic speed-limit expressions derived in this work are exact within the effective model introduced in Secs.~\ref{sec:QSL-two-mode}-\ref{sec:relativistic}. Their interpretation for concrete optical-link architectures depends on additional assumptions about where the relativistic shift enters, how the local oscillator is generated, and how the heterodyne receiver is implemented.

{
Section~\ref{sec:time-resolution} showed that $\Psi_\star(t_0)\to0$ in the
bright-LO regime when the detector-frame beat frequency is bounded. The
operational question is therefore the temporal precision of the demodulated
intermediate-frequency phase record. We use the distinction among the
joint-state QSL time $t_\star$, the accumulated beat phase $\Psi_\star(t_0)$,
and the finite-window readout uncertainty $\Delta\psi(T)$ to formulate
detector-front-end timing benchmarks. Within the present model, the leading
local benchmark is set by the total demodulated phase-noise ASD
$S_\psi^{1/2}$ together with the operating-point beat note $\Delta\omega'_0$,
which depends on the baseline beat note and the local kinematic factor
$\kappa(t_0)$. Link-specific use of this benchmark also requires the residual
tracking convention and local chirp rate.
}

{
At a chosen detector-frame operating epoch $t_0$, the local frequency rescaling
$\omega_{\mathrm{LO}}'(t_0)=\kappa(t_0)\omega_{\mathrm{LO}}$ gives the beat note
\begin{equation}
\Delta\omega'_0
\equiv
\Delta\omega'(t_0)
=
\Delta\omega_0-\bigl[\kappa(t_0)-1\bigr]\omega_{\mathrm{LO}} .
\label{eq:Deltaomega_prime_main}
\end{equation}
Here $\Delta\omega_0\equiv\omega_s-\omega_{\rm LO}$ is the unshifted local
baseline beat note. When no time argument is displayed below, $\Delta\omega'$
denotes this local operating-point value. Time-dependent beat phases are
governed by the accumulated phase relation introduced in
Sec.~\ref{sec:time-resolution}; this is the appropriate representation whenever
Doppler, gravitational, tracking-reference, or demodulation-reference
contributions vary across the observation window
\cite{Turyshev2014PRD,BayleHartwigStaab2021PRD}.
}

It is important to distinguish the target metrology: continuous-wave optical ranging is set by the optical phase itself through
$
\Delta L=(\lambda/2\pi)\Delta\psi=(c/\omega_s)\Delta\psi
$
and is therefore not directly degraded by a change in the heterodyne beat frequency. By contrast, absolute time-transfer, clock synchronization, and related temporal tasks are tied to the intermediate-frequency (IF) phase readout produced by the phasemeter front end, which in practice is obtained from digital phase estimation and loop-based tracking of the electrical beat note rather than from literal zero-crossing counting \cite{Shaddock_2006,Bachman2017GRACEFO}.
{
For timing-estimation comparisons in which the finite-window phase uncertainty
$\Delta\psi(T)$ is held fixed for the comparison, the local phase-to-time
conversion gives
\begin{equation}
\Delta t_{\mathrm{het}}(T)
=
\frac{\Delta\psi(T)}{|\Delta\omega_0|},
\qquad
\Delta t_{\mathrm{het}}'(T)
=
\frac{\Delta\psi(T)}{|\Delta\omega'_0|}.
\label{eq:dt_from_dpsi_prime}
\end{equation}
This phase-to-time conversion uses the locally stationary phase map over the
same averaging window $T$. Equivalently, it assumes that the chirp correction is
negligible on that window, for example
\begin{equation}
|\dot{\Delta\omega}'_0|\,T \ll |\Delta\omega'_0| .
\label{eq:local_stationarity_condition_main}
\end{equation}
Hence the detector-frame rescaling implies
\begin{equation}
\Delta t_{\mathrm{het}}'(T)
=
\Delta t_{\mathrm{het}}(T)\,
\frac{|\Delta\omega_0|}{|\Delta\omega'_0|}.
\label{eq:dt_scaling_main}
\end{equation}
This scaling isolates only the kinematic phase-to-time conversion. If the
demodulated phase-noise ASD itself depends on heterodyne frequency or receiver
bandwidth, that dependence must be included separately in $\Delta\psi(T)$ or
$S_\psi$.
}

At standard optical wavelengths ($\lambda=1064~\mathrm{nm}$), the carrier satisfies $\omega_{\mathrm{LO}}\sim 10^{15}~\mathrm{rad/s}$. {Small fractional frame shifts therefore induce large absolute local beat-note shifts}
{
\begin{equation}
\delta f_0
=
|\Delta f'_0-\Delta f_0| .
\label{eq:deltaf_def}
\end{equation}
}
{
Assuming the frequency plan is chosen so that $|\Delta f'_0|$ remains within the electrical front-end bandwidth of the photodiode, these local shifts consume front-end frequency margin. Their magnitudes across representative operating points are summarized in Table~\ref{tab:carrier2beat}.
}

\begin{table}[htbp]
\caption{{Representative instantaneous carrier-to-beat amplification at selected operating points for an optical carrier at $\lambda=1064~\mathrm{nm}$. The entries are local detector-frame benchmarks evaluated at a chosen epoch, rather than complete time-dependent orbital or tracking models. LISA row reflects actively minimized relative line-of-sight kinematics between constellation spacecraft, whereas the Deep Space  row reflects raw heliocentric orbital drift.}}
\label{tab:carrier2beat}
\begin{ruledtabular}
\begin{tabular}{lccccc}
System & Relative $\beta_\parallel$ & $\Delta\Phi/c^2$ & $\Delta f_0$ & {Local shift $|\delta f_0|$} & {$|\Delta f'_0|$} \\
\hline
Deep Space Link & $1.0\times 10^{-4}$ & $1.0\times 10^{-8}$ & $5.0~\mathrm{GHz}$  & $28.2~\mathrm{GHz}$ & $23.2~\mathrm{GHz}$ \\
LEO downlink    & $2.502\times 10^{-5}$ & $4.0\times 10^{-11}$ & $10.0~\mathrm{GHz}$ & $7.05~\mathrm{GHz}$ & $2.95~\mathrm{GHz}$ \\
LISA            & $5.0\times 10^{-8}$ & $\sim 0$            & $20.0~\mathrm{MHz}$ & $14.1~\mathrm{MHz}$ & $5.9~\mathrm{MHz}$ \\
\end{tabular}
\end{ruledtabular}
\end{table}

{
The table supplies the instantaneous values entering the local phase-to-time map. Applying Eqs.~\eqref{eq:local_stationarity_condition_main} and \eqref{eq:chirp_small_parameter_main} to a specific link additionally requires the local chirp rate determined by the orbit, frequency plan, and tracking convention.
}

{
A finite acquisition window depends on the residual beat phase that remains after the chosen prediction, preprocessing, or demodulation reference has been applied.
} {
Consider an untracked, open-loop integration window during transient acquisition before active phase-locked loop (PLL) engagement. We assume that the demodulated phase record lies on a locally white one-sided phase-noise plateau.  With the rectangular-window convention introduced in Sec.~\ref{sec:time-resolution}, the stochastic contribution to the averaged phase estimator is $\mathrm{Var}(\hat\psi_T)=S_\psi/(2T)$ \cite{BendatPiersol2010,Saulson2017}. The deterministic contribution is the residual phase accumulated from the remaining beat-frequency mismatch. Let $\delta\Omega(u)$ denote the residual detector-frame beat-frequency mismatch after any open-loop Doppler prediction, preprocessing, or demodulation reference has been applied. Across the averaging window, the residual phase then satisfies
\begin{equation}
\phi(T)-\phi(0)
=
\int_0^T \delta\Omega(u)\,du .
\label{eq:general_residual_phase_main}
\end{equation}
For a uniform window-average phase estimator referenced to the initial phase, the deterministic bias is
\begin{equation}
b(T)
=
\frac{1}{T}\int_0^T[\phi(t)-\phi(0)]\,dt
=
\frac{1}{T}\int_0^T(T-u)\delta\Omega(u)\,du .
\label{eq:general_bias_functional_main}
\end{equation}
Thus the open-loop phase MSE in the white-noise plateau is
\begin{equation}
\mathrm{MSE}(T)
=
\frac{S_\psi}{2T}
+
\left[
\frac{1}{T}\int_0^T(T-u)\delta\Omega(u)\,du
\right]^2 .
\label{eq:general_MSE_functional_main}
\end{equation}
Equation~\eqref{eq:general_MSE_functional_main} is therefore the appropriate open-loop MSE functional for a general time-dependent residual beat phase.
}

{
The cube-root benchmark is obtained in the nonzero constant-mismatch, locally
quasi-static limit of Eq.~\eqref{eq:general_MSE_functional_main}. Expanding the
residual mismatch about the operating epoch,
\begin{equation}
\delta\Omega(u)
=
\delta\Omega_0+\dot{\delta\Omega}_0u+O(u^2),
\label{eq:residual_mismatch_expansion_main}
\end{equation}
gives
\begin{equation}
b(T)
=
\frac{1}{2}\delta\Omega_0T
+
\frac{1}{6}\dot{\delta\Omega}_0T^2
+
O(T^3).
\label{eq:bias_expansion_main}
\end{equation}
Thus the constant-mismatch approximation is controlled by
\begin{equation}
\epsilon_{\rm chirp}(T)
\equiv
\frac{|\dot{\delta\Omega}_0|\,T}{3|\delta\Omega_0|}
\ll 1 .
\label{eq:chirp_small_parameter_main}
\end{equation}
}

{
In a realistic receiver, $\delta\Omega_0$ denotes the residual mismatch after
the chosen acquisition, Doppler-estimation, tracking, or preprocessing
convention. Identifying $\delta\Omega_0$ with the raw table shift is therefore a
deliberately untracked open-loop benchmark, not a model of the residual mismatch
in a locked receiver \cite{Paillier2020JLT,McSorley2025PRApplied,Valliyakalayil2022PRD}.
When this open-loop benchmark is intended and
$\delta\Omega_0\equiv\delta\omega=2\pi\delta f_0\neq0$,
Eq.~\eqref{eq:general_MSE_functional_main} reduces to
\begin{equation}
\mathrm{MSE}_{\rm qs}(T)
=
\frac{S_\psi}{2T}
+
\frac{1}{4}\bigl(\delta\omega\,T\bigr)^2 .
\label{eq:MSE_quasistatic_main}
\end{equation}
Minimizing this quasi-static metric yields
\begin{equation}
T_{\mathrm{opt}}^{\rm qs}
=
\left(\frac{S_\psi^{1/2}}{|\delta\omega|}\right)^{2/3}.
\label{eq:Topt_quasistatic_main}
\end{equation}
At this optimum, the minimum MSE is
\begin{equation}
\mathrm{MSE}_{\min}^{\rm qs}
=
\frac{3}{4}\,S_\psi^{2/3}\,|\delta\omega|^{2/3},
\qquad
\sqrt{\mathrm{MSE}_{\min}^{\rm qs}}
=
\sqrt{\frac{3}{4}}\,S_\psi^{1/3}\,|\delta\omega|^{1/3}.
\label{eq:MSEmin_quasistatic_main}
\end{equation}
}

{
If $\epsilon_{\rm chirp}(T_{\mathrm{opt}}^{\rm qs})$ is of order unity, the
deterministic bias must be evaluated from
Eq.~\eqref{eq:general_bias_functional_main}, and the cube-root scaling need not
apply. For LEO optical links, GHz-scale Doppler shifts accompanied by large
Doppler rates can violate the quasi-static condition over long acquisition
windows \cite{McSorley2025PRApplied}. The quasi-static reduction is therefore
restricted to transient-acquisition windows satisfying
Eq.~\eqref{eq:chirp_small_parameter_main}. For the LISA and deep-space rows in
Table~\ref{tab:carrier2beat}, the local chirp rate $\dot{\delta\Omega}_0$ is
not specified by the instantaneous shift entries. For deep-space links, the
finite-window approximation is trajectory dependent: the local Doppler scale in
Table~\ref{tab:carrier2beat} must be supplemented by the local Doppler rate and,
when needed, higher derivatives along the chosen ephemeris.
}

\begin{figure}[htbp]
\centering
\includegraphics[width=0.9\linewidth]{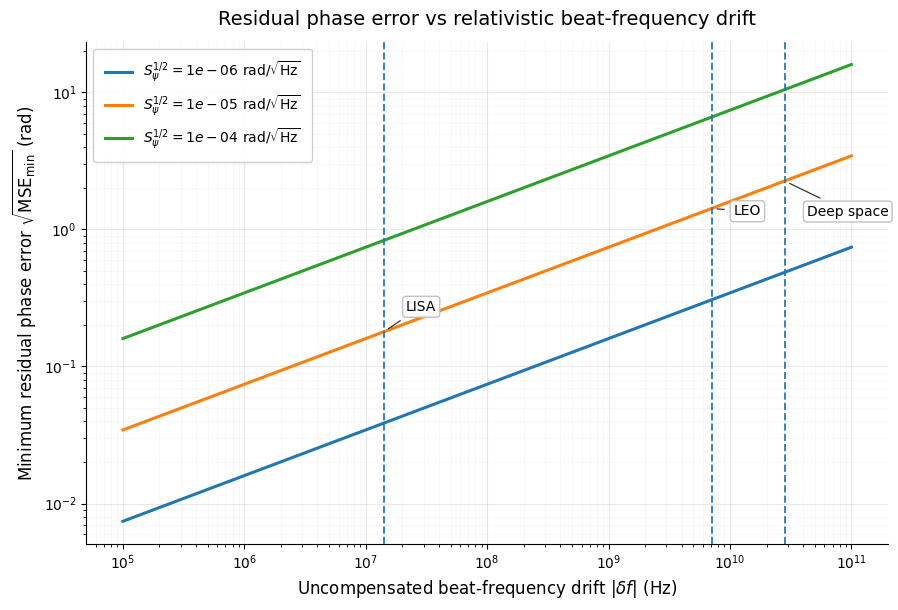}
\caption{{Quasi-static open-loop benchmark for the minimum residual heterodyne phase error, $\sqrt{\mathrm{MSE}_{\min}^{\rm qs}}$, versus local uncompensated beat-frequency offset, $|\delta f_0|$, for three total demodulated white phase-noise ASD levels, $S_{\psi}^{1/2}$. The curves assume a locally constant residual beat-frequency mismatch $\delta\omega=2\pi\delta f_0$ over the optimized averaging window, equivalently $\epsilon_{\rm chirp}(T_{\mathrm{opt}}^{\rm qs})\ll1$. The dashed vertical lines mark representative local offset values for LISA, LEO, and deep-space links from Table~\ref{tab:carrier2beat}.}}
\label{fig:cuberoot_phase_error}
\end{figure}

{
Figure~\ref{fig:cuberoot_phase_error} displays the quasi-static cube-root
benchmark on log--log axes. The fixed-$S_\psi^{1/2}$ curves have slope $1/3$,
and the dashed vertical markers place the representative LISA, LEO, and
deep-space operating points from Table~\ref{tab:carrier2beat} on the same
detector-frame scale. The figure should therefore be read as an untracked
open-loop benchmark within the quasi-static regime, not as a closed-loop
phasemeter performance curve.
}

{
The quasi-static optimum is sub-nanosecond when
\begin{equation}
S_\psi^{1/2}
<
2\pi |\delta f_0|\,(1~\mathrm{ns})^{3/2}.
\label{eq:subnanosecond_ASD_threshold}
\end{equation}
For the representative local shifts in Table~\ref{tab:carrier2beat}, this gives
\begin{equation}
\begin{aligned}
S_\psi^{1/2} &< 2.80~\mu\mathrm{rad}/\sqrt{\mathrm{Hz}}
&& \text{for LISA}, \\
S_\psi^{1/2} &< 1.40~\mathrm{mrad}/\sqrt{\mathrm{Hz}}
&& \text{for LEO}, \\
S_\psi^{1/2} &< 5.60~\mathrm{mrad}/\sqrt{\mathrm{Hz}}
&& \text{for the deep-space benchmark}.
\end{aligned}
\end{equation}
These thresholds follow from Eq.~\eqref{eq:Topt_quasistatic_main} and the table values. Whether an actual mission realizes these thresholds depends on the total demodulated phase-noise ASD of the complete receiver; in space-interferometric readout, phasemeter noise is only one contribution among laser-frequency noise, shot noise, relative-intensity noise, optical-path noise, and pointing or tilt-to-length couplings \cite{Gerberding2015RSI,Armano2021LPFOMS}.
}

{
For representative low-noise phase-ASD values satisfying Eq.~\eqref{eq:subnanosecond_ASD_threshold}, large uncompensated detector-frame beat-frequency offsets can push the quasi-static $T_{\mathrm{opt}}^{\rm qs}$ into the sub-nanosecond regime, corresponding to an inverse time scale above $1~\mathrm{GHz}$. This inverse acquisition-time scale is well beyond the tens-of-MHz heterodyne bands and photoreceiver bandwidths used in LISA-like space-interferometry implementations, where prototype photoreceivers report 3-dB bandwidths of order $34~\mathrm{MHz}$ \cite{Guzman_2011,HeinzelFrequencyPlanning2024}. Constellation architectures like LISA therefore require offset-frequency planning and active locking to keep Doppler-shifted beat notes within the phasemeter heterodyne band \cite{Shaddock_2006,HeinzelFrequencyPlanning2024}. A properly acquired closed-loop PLL tracks constant or slowly varying frequency offsets within its capture range and bandwidth. The residual quasi-static window $T_{\mathrm{opt}}^{\rm qs}$ therefore characterizes transient acquisition, loss-of-lock recovery, or cycle-slip recovery, not the generic steady-state limit of a locked phasemeter \cite{Shaddock_2006}.
}

{
After separating acquisition-bandwidth constraints from steady-state phasemeter
tracking, the remaining question is whether quantum-noise suppression can compensate
the detector-frame beat-note rescaling in the reduced heterodyne model.
}

Continuous-variable squeezing can mitigate phase noise only after the squeezing resource is specified consistently with the heterodyne measurement model. For phase-insensitive heterodyne detection, the measured phase-noise floor receives contributions from both the signal band and the image band, so squeezing only the signal band does not by itself remove the full image-band penalty \cite{Anai2024Heterodyne}. A quantitatively complete phase-insensitive heterodyne model must also include the image-band vacuum contribution and the associated receiver architecture. Appendix~\ref{app:sqcoh-BHD} gives the corresponding reduced single-mode bound and shows how the required suppression factor enters the break-even estimate. In the reduced single-mode benchmark used here, the bright-LO phase uncertainty acquires the effective suppression factor $e^{-r}$. Imposing the break-even condition
{
\begin{equation}
\Delta t_{\mathrm{het}}^{(\mathrm{sq})}(T;\Delta\omega'_0)
=
\Delta t_{\mathrm{het}}^{(\mathrm{coh})}(T;\Delta\omega_0)
\label{eq:breakeven_condition_main}
\end{equation}
}
together with Eq.~\eqref{eq:dt_from_dpsi_prime} then yields the required squeezing level:
{
\begin{equation}
r_{\mathrm{req}}
=
\max\!\left(0,\,
-\ln\left|1-\bigl[\kappa(t_0)-1\bigr]\frac{\omega_{\mathrm{LO}}}{\Delta\omega_0}\right|
\right).
\label{eq:breakeven_scaling_relation}
\end{equation}
Equivalently,
\[
r_{\mathrm{req}}
=
\max\!\left(0,-\ln\left|\frac{\Delta\omega'_0}{\Delta\omega_0}\right|\right).
\]
}

{
Within the local IF phase-to-time map used in Eq.~\eqref{eq:dt_from_dpsi_prime}, driving the local absolute beat note to zero, $|\Delta\omega'_0|\to0$, makes the linear timing conversion singular because the local phase derivative $|d\psi/dt|_{t_0}$ vanishes. In LISA-like DPLL phasemeters, beat notes approaching zero frequency are explicitly excluded because the digital phase-locked loop can lose lock near this limit \cite{HeinzelFrequencyPlanning2024}. This limitation concerns the present heterodyne timing estimator, not all near-DC or nonlinear phase-tracking architectures. Dual-quadrature phasemeter schemes, for example, have been proposed for operation
down to DC \cite{SambridgeMcKenzie2025PRApplied}.
}

{
Consider a representative low-Earth-orbit downlink initialized at
$\Delta f_0=10.0~\mathrm{GHz}$. During a maximal approaching line-of-sight pass
($v_\parallel\simeq 7.5~\mathrm{km/s}$, $\beta_\parallel\simeq 2.502\times10^{-5}$),
the optical carrier blueshifts by $\approx7.05~\mathrm{GHz}$
(Table~\ref{tab:carrier2beat}), compressing the available heterodyne offset to
$|\Delta f'_0|\simeq2.95~\mathrm{GHz}$. Substituting this local
operating-point beat-note ratio into
Eq.~\eqref{eq:breakeven_scaling_relation} gives
$r_{\mathrm{req}}\simeq1.22~\mathrm{Np}$, or about $10.6~\mathrm{dB}$ of
detected quantum squeezing, comparable to high-level continuous-variable
squeezing demonstrations \cite{Vahlbruch2016}.
}

{
Because the channel efficiency $\eta\in(0,1]$ maps the injected source state to
the detected variance through the passive loss model
\cite{Weedbrook2012,BachorRalph2004},
\begin{equation}
V_{\mathrm{det}}
=
1-\eta+\eta e^{-2r_{\mathrm{src}}},
\label{eq:passive_loss_variance}
\end{equation}
the target $V_{\mathrm{det}}\le10^{-10.6/10}\approx0.087$ imposes an effective
transmissivity floor $\eta\ge0.913$, reached only in the limit of arbitrarily
strong injected source squeezing. This threshold is therefore a model-dependent
benchmark, not a universal no-go bound for all heterodyne architectures.
}

{
Atmospheric LEO downlinks typically have transmissivities far below the floor
implied by this threshold. For example, the Micius satellite-to-ground experiment
estimated about $22~\mathrm{dB}$ diffraction loss at $1200~\mathrm{km}$, with an
additional $3$--$8~\mathrm{dB}$ from atmospheric absorption and turbulence and
less than $3~\mathrm{dB}$ from pointing error \cite{Liao2017}. Such links therefore
fall far below the transmissivity needed to recover the baseline timing-resolution
benchmark within the reduced single-mode heterodyne resource model, even in the
limit of arbitrarily strong injected source squeezing. This identifies very
low-loss vacuum or near-vacuum channels, such as intersatellite links, as the most
plausible regime of applicability for the present reduced model.
}


\section{Conclusion}
\label{sec:conclusion_main}
We have developed a detector-frame quantum-speed-limit framework for balanced heterodyne readout within an effective signal--local-oscillator two-mode model, where relativistic motion enters as a kinematic rescaling of the local-oscillator frequency. The main conceptual result is the appearance of a geometric--operational mismatch in the bright-local-oscillator regime: even though the full detector-frame two-mode quantum state becomes rapidly distinguishable under the joint Hamiltonian dynamics, the practically relevant limitation remains the task-restricted temporal precision of the intermediate-frequency phase record obtained from the heterodyne measurement channel. In this regime, the operational information rate relevant for phase tracking is therefore determined by the measurement process, rather than by the global distinguishability structure of the underlying optical state.

At the operational level, we studied temporal phase estimation in an untracked open-loop transient-acquisition model before active phase locking. By balancing stochastic white phase noise against deterministic residual beat-frequency drift, we obtained an optimal pre-processed integration window and a cube-root drift scaling relation for the minimum residual phase error. This scaling gives a direct quantitative link between the detector-frame state-geometry layer described by quantum speed limits and the practical estimation problem defined by the heterodyne phase record. Within the reduced single-mode Gaussian resource-scaling description used here, the same timing model also gives a squeezing--kinematic break-even condition for temporal intermediate-frequency readout tasks. For a representative low-Earth-orbit scenario, assuming a compressive frequency plan, this condition corresponds to an effective requirement of about $10.6$~dB of detected continuous-variable squeezing, together with an absolute best-case transmissivity floor of $\eta \ge 0.913$, reached only in the limit of arbitrarily strong injected squeezing.

Within the present temporal heterodyne readout model, these results therefore serve as quantitative detector-frame benchmarks and point to very low-loss vacuum or near-vacuum optical channels---such as intersatellite links---as the natural regime of applicability. The analysis also identifies a limiting configuration in which the relativistically shifted beat note approaches zero and the measurement is pushed toward a DC homodyne regime, where the time-to-phase conversion becomes singular and continuous timing information is lost within the heterodyne framework.

More broadly, the present results show how global dynamical limits derived from quantum speed limits can decouple from the operational limits that govern continuous quantum measurements. Rather than giving a universal no-go theorem for relativistic heterodyne phase tracking, the analysis identifies a controlled class of Gaussian effective models in which global joint-state distinguishability becomes sharply separated from task-restricted measurement performance. In this way, the framework developed here provides a controlled benchmark for relativistic heterodyne timing in bright-reference Gaussian architectures and clarifies how quantum speed-limit geometry is related to practical phase-tracking limits in continuous-variable optical metrology.

\bibliography{references}

\appendix

\section{Quantum Speed Limit for Two-Mode Product States}\label{app:two-mode-QSL}

We consider two independent harmonic modes, signal $s$ and local oscillator $\mathrm{LO}$, with annihilation operators
$\hat a_s,\hat a_{\mathrm{LO}}$ and respective frequencies $\omega_s,\omega_{\mathrm{LO}}$.
The free Hamiltonian is
\begin{equation}
\hat{H}
= \hbar\,\omega_{s}\Bigl(\hat{a}_{s}^\dagger \hat{a}_{s}+\tfrac{1}{2}\Bigr)
+ \hbar\,\omega_{\mathrm{LO}}\Bigl(\hat{a}_{\mathrm{LO}}^\dagger \hat{a}_{\mathrm{LO}}+\tfrac{1}{2}\Bigr).
\label{eq:app_Hfree}
\end{equation}
We study three two-mode product inputs:
(i) coherent$\otimes$coherent,
(ii) squeezed-vacuum$\otimes$coherent,
(iii) squeezed-coherent$\otimes$coherent.

For a pure initial product state $\rho_0=\ket{\psi_0}\!\bra{\psi_0}$ evolving under $\hat H$,
the fidelity at time $T$ is
\begin{equation}
F(T)=\bigl|\langle \psi_0\vert e^{-i\hat H T/\hbar}\vert \psi_0\rangle\bigr|^{2}.
\end{equation}
The Bures angle is $\mathcal{L}(T)=\arccos(\sqrt{F(T)})$.
We define the ground energy
\begin{equation}
E_0=\frac{\hbar}{2}\,(\omega_s+\omega_{\mathrm{LO}}),
\end{equation}
together with the mean energy and energy variance in the initial state,
\begin{equation}
\langle H\rangle=\langle \psi_0\vert \hat H\vert\psi_0\rangle,\qquad
(\Delta H)^2=\langle \psi_0\vert \hat H^2\vert\psi_0\rangle-\langle H\rangle^2.
\end{equation}
We use the standard Mandelstam-Tamm and Margolus-Levitin forms (with explicit ground-energy subtraction):
\begin{equation}
\mathcal{T}_{\mathrm{MT}}(T)=\frac{\hbar\,\mathcal{L}(T)}{\Delta H},\qquad
\mathcal{T}_{\mathrm{ML}}(T)=\frac{\hbar}{\langle H\rangle-E_0}\,\bigl(1-\cos\mathcal{L}(T)\bigr),
\label{eq:app_TMT_TML}
\end{equation}
and the QSL time at horizon $T$ is
\begin{equation}
\tau_{\mathrm{QSL}}(T)=\max\{\mathcal{T}_{\mathrm{MT}}(T),\,\mathcal{T}_{\mathrm{ML}}(T)\}.
\label{eq:app_tauQSL}
\end{equation}
The QSL expressions are understood for non-stationary inputs; for exact energy eigenstates the fidelity is time-independent and the corresponding evolution is trivial. Because the two modes evolve independently, all fidelities factorize into a signal factor times an LO factor.

\paragraph{Case (i): coherent$\otimes$coherent.}
Let $\ket{\psi_0}=\ket{\alpha_s}\otimes\ket{\alpha_{\mathrm{LO}}}$.
Free evolution rotates amplitudes,
$\alpha_s\mapsto \alpha_s e^{-i\omega_s T}$, $\alpha_{\mathrm{LO}}\mapsto \alpha_{\mathrm{LO}}e^{-i\omega_{\mathrm{LO}}T}$.
The single-mode coherent overlap yields
\begin{equation}
\bigl|\langle \alpha \vert \alpha e^{-i\omega T}\rangle\bigr|^2
=\exp\!\Bigl[-4|\alpha|^{2}\sin^{2}\!\Bigl(\frac{\omega T}{2}\Bigr)\Bigr],
\end{equation}
hence the two-mode fidelity is
\begin{equation}
F_{\text{cc}}(T)
=\exp\!\Bigl[-4|\alpha_s|^{2}\sin^{2}\!\Bigl(\frac{\omega_s T}{2}\Bigr)\Bigr]\,
 \exp\!\Bigl[-4|\alpha_{\mathrm{LO}}|^{2}\sin^{2}\!\Bigl(\frac{\omega_{\mathrm{LO}} T}{2}\Bigr)\Bigr].
\label{eq:app_F_cc}
\end{equation}
The mean energy and variance follow from $\hat H=\hbar\omega_s(\hat n_s+\tfrac{1}{2})+\hbar\omega_{\mathrm{LO}}(\hat n_{\mathrm{LO}}+\tfrac{1}{2})$ with $\langle \hat n\rangle=|\alpha|^2$, $\mathrm{Var}(\hat n)=|\alpha|^2$ for coherent states:
\begin{align}
\langle H\rangle_{\text{cc}}
&=\hbar\omega_s\Bigl(|\alpha_s|^2+\tfrac{1}{2}\Bigr)+\hbar\omega_{\mathrm{LO}}\Bigl(|\alpha_{\mathrm{LO}}|^2+\tfrac{1}{2}\Bigr),\\
(\Delta H)^2_{\text{cc}}
&=\hbar^2\Bigl(\omega_s^2|\alpha_s|^2+\omega_{\mathrm{LO}}^2|\alpha_{\mathrm{LO}}|^2\Bigr).
\end{align}
Insert \eqref{eq:app_F_cc} into $\mathcal{L}(T)$ and then into \eqref{eq:app_TMT_TML}.

\paragraph{Case (ii): squeezed-vacuum$\otimes$coherent.}
Let the signal be squeezed vacuum $\ket{\mathrm{SqVac}(r,0)}:=\hat S(r)\ket{0}$ with $r\ge 0$ and $\hat S(r)\equiv \hat S(\zeta)$ at $\zeta=r$ (squeezing phase $0$).
Let the LO be $\ket{\alpha_{\mathrm{LO}}}$, so $\ket{\psi_0}=\ket{\mathrm{SqVac}(r,0)}\otimes\ket{\alpha_{\mathrm{LO}}}$.

Under free evolution, the squeezing phase advances by $2\omega_s T$.
The overlap between two squeezed vacua with equal squeezing strength $r$ and relative squeezing-phase shift $2\omega_s T$ is
\begin{equation}
\langle \mathrm{SqVac}(r,0)\vert \mathrm{SqVac}(r,2\omega_s T)\rangle
=\Bigl(\cosh^2 r - e^{i\,2\omega_s T}\sinh^2 r\Bigr)^{-1/2}.
\end{equation}
Taking modulus squared gives the signal fidelity factor
\begin{equation}
F_{\mathrm{sv}}(T)
=\frac{1}{\bigl|\cosh^2 r - e^{i\,2\omega_s T}\sinh^2 r\bigr|}
=\frac{1}{\sqrt{1+\sinh^2(2r)\,\sin^2(\omega_s T)}}.
\label{eq:app_F_sv_signal}
\end{equation}
The LO coherent factor is the same as in case (i), so the two-mode fidelity is
\begin{equation}
F_{\text{sv}\otimes\text{c}}(T)
=F_{\mathrm{sv}}(T)\;
 \exp\!\Bigl[-4|\alpha_{\mathrm{LO}}|^{2}\sin^{2}\!\Bigl(\frac{\omega_{\mathrm{LO}} T}{2}\Bigr)\Bigr].
\label{eq:app_F_svxc}
\end{equation}
Energy moments use $\langle \hat n_s\rangle=\sinh^2 r$ and
$\mathrm{Var}(\hat n_s)=2\sinh^2 r\,\cosh^2 r$ for squeezed vacuum:
\begin{align}
\langle H\rangle_{\text{sv}\otimes\text{c}}
&=\hbar\omega_s\Bigl(\sinh^2 r+\tfrac{1}{2}\Bigr)+\hbar\omega_{\mathrm{LO}}\Bigl(|\alpha_{\mathrm{LO}}|^2+\tfrac{1}{2}\Bigr),\\
(\Delta H)^2_{\text{sv}\otimes\text{c}}
&=\hbar^2\Bigl(\omega_s^2\,2\sinh^2 r\,\cosh^2 r+\omega_{\mathrm{LO}}^2|\alpha_{\mathrm{LO}}|^2\Bigr).
\label{eq:app_DH2_svxc}
\end{align}
Insert \eqref{eq:app_F_svxc} into $\mathcal{L}(T)$, then apply \eqref{eq:app_TMT_TML}.

\paragraph{Case (iii): squeezed-coherent$\otimes$coherent.}
In the main text, this single-mode squeezed-coherent family is used only as a reduced resource-scaling model for the temporal readout discussion in Sec.~\ref{sec:applications_main}; a full phase-insensitive heterodyne implementation requires an explicit signal/image-band treatment.
We use the displaced squeezed vacuum convention
\begin{equation}
\ket{\alpha_s,r}:=\hat D(\alpha_s)\hat S(r)\ket{0},\qquad
\ket{\psi_0}=\ket{\alpha_s,r}\otimes\ket{\alpha_{\mathrm{LO}}}.
\label{eq:app_def_sqcoh}
\end{equation}
Free evolution rotates both displacement and squeezing axis in phase space.

\emph{Signal fidelity factor in closed form.}
Introduce quadratures $\hat x=(\hat a+\hat a^\dagger)/\sqrt2$, $\hat p=(\hat a-\hat a^\dagger)/(i\sqrt2)$, and the phase-space rotation
\begin{equation}
R(\varphi)=\begin{pmatrix}\cos\varphi&-\sin\varphi\\ \sin\varphi&\cos\varphi\end{pmatrix},\qquad \varphi=\omega_s T.
\end{equation}
For $\ket{\alpha_s,r}$ with squeezing axis along $\hat x$ at $T=0$, the covariance matrix and displacement vector are
\begin{equation}
V_0=\frac{1}{2}\begin{pmatrix}e^{-2r}&0\\0&e^{2r}\end{pmatrix},\qquad
d_0=\begin{pmatrix}\sqrt2\,\mathrm{Re}\,\alpha_s\\ \sqrt2\,\mathrm{Im}\,\alpha_s\end{pmatrix}.
\end{equation}
At time $T$, these evolve as $V_T=R(\varphi)V_0R(\varphi)^{\mathsf{T}}$ and $d_T=R(\varphi)d_0$.
For two pure single-mode Gaussian states $(V_0,d_0)$ and $(V_T,d_T)$, the fidelity is
\begin{equation}
F_{\mathrm{sc}}^{(s)}(T)
=\frac{\exp\!\Bigl[-\frac{1}{2}\,\delta d^{\mathsf{T}}\,(V_0+V_T)^{-1}\,\delta d\Bigr]}{\sqrt{\det(V_0+V_T)}},
\qquad \delta d:=d_0-d_T.
\label{eq:app_F_sc_signal_general}
\end{equation}
Writing $c=\cos\varphi$, $s=\sin\varphi$, the matrix elements of $V_0+V_T$ are
\begin{align}
(V_0+V_T)_{11}&=\frac{1}{2}\Bigl[e^{-2r}(1+c^2)+e^{2r}s^2\Bigr],\\
(V_0+V_T)_{22}&=\frac{1}{2}\Bigl[e^{2r}(1+c^2)+e^{-2r}s^2\Bigr],\\
(V_0+V_T)_{12}&=\frac{1}{2}\,(e^{-2r}-e^{2r})\,cs,
\end{align}
with determinant
\begin{equation}
\det(V_0+V_T)=1+\sinh^2(2r)\,\sin^2(\varphi).
\label{eq:app_det_Vsum}
\end{equation}
Equation \eqref{eq:app_F_sc_signal_general} together with \eqref{eq:app_det_Vsum} gives the signal fidelity in closed form.

\emph{Total two-mode fidelity.}
The LO coherent factor matches case (i), hence
\begin{equation}
F_{\text{sc}\otimes\text{c}}(T)
=F_{\mathrm{sc}}^{(s)}(T)\;
 \exp\!\Bigl[-4|\alpha_{\mathrm{LO}}|^{2}\sin^{2}\!\Bigl(\frac{\omega_{\mathrm{LO}} T}{2}\Bigr)\Bigr].
\label{eq:app_F_scxc}
\end{equation}

\emph{Energy moments.}
Let $N=\sinh^2 r$ and $M=-\frac{1}{2}\sinh(2r)$ (squeezing phase $0$).
For the signal mode in $\ket{\alpha_s,r}$,
\begin{equation}
\langle \hat n_s\rangle=|\alpha_s|^2+N,
\qquad
\mathrm{Var}(\hat n_s)=|\alpha_s|^2\cosh(2r)-\mathrm{Re}(\alpha_s^2)\sinh(2r)+2N(N+1).
\label{eq:app_var_n_sqcoh}
\end{equation}
The LO mode is coherent with $\langle \hat n_{\mathrm{LO}}\rangle=|\alpha_{\mathrm{LO}}|^2$ and $\mathrm{Var}(\hat n_{\mathrm{LO}})=|\alpha_{\mathrm{LO}}|^2$.
Therefore,
\begin{align}
\langle H\rangle_{\text{sc}\otimes\text{c}}
&=\hbar\omega_s\Bigl(|\alpha_s|^2+N+\tfrac{1}{2}\Bigr)+\hbar\omega_{\mathrm{LO}}\Bigl(|\alpha_{\mathrm{LO}}|^2+\tfrac{1}{2}\Bigr),\\
(\Delta H)^2_{\text{sc}\otimes\text{c}}
&=\hbar^2\Bigl(\omega_s^2\,\mathrm{Var}(\hat n_s)+\omega_{\mathrm{LO}}^2|\alpha_{\mathrm{LO}}|^2\Bigr),
\end{align}
with $\mathrm{Var}(\hat n_s)$ from \eqref{eq:app_var_n_sqcoh}.
Insert \eqref{eq:app_F_scxc} into $\mathcal{L}(T)$ and then apply \eqref{eq:app_TMT_TML} and \eqref{eq:app_tauQSL}.

\section{Relativistic Frequency Shifts and Quantum Speed Limits}\label{app:relativistic-shift}

A distant observer may infer a shifted LO frequency of the form
\begin{equation}
\omega_{\mathrm{LO}}'
=\kappa\,\omega_{\mathrm{LO}},
\qquad
\kappa = 1+\beta_\parallel+\frac{\Delta\Phi}{c^2},
\label{eq:app_kappa_def}
\end{equation}
where $\beta_\parallel$ parameterizes the leading longitudinal Doppler effect and $\Delta\Phi/c^2$ represents the leading weak-field gravitational potential difference between the emission and reception points. Throughout this appendix, the shifted Hamiltonian is understood as the effective detector-frame single-mode model introduced in Sec.~\ref{sec:relativistic}. Replacing $\omega_{\mathrm{LO}}\to\omega_{\mathrm{LO}}'$ yields the observer-frame Hamiltonian
\begin{equation}
\hat H'
=\hbar\omega_s\Bigl(\hat a_s^\dagger\hat a_s+\tfrac{1}{2}\Bigr)
+\hbar\omega_{\mathrm{LO}}'\Bigl(\hat a_{\mathrm{LO}}^\dagger\hat a_{\mathrm{LO}}+\tfrac{1}{2}\Bigr).
\end{equation}
All QSL ingredients follow by substituting $\omega_{\mathrm{LO}}\to\omega_{\mathrm{LO}}'$ in Appendix~\ref{app:two-mode-QSL}.

The ground energy becomes
\begin{equation}
E_0'=\frac{\hbar}{2}\,(\omega_s+\omega_{\mathrm{LO}}'),
\end{equation}
and the QSL functions keep the form \eqref{eq:app_TMT_TML} with $\langle H'\rangle$, $\Delta H'$, $E_0'$, and the shifted LO overlap.

\paragraph{Coherent$\otimes$coherent.}
\begin{align}
F_{\text{cc}}'(T)
&=\exp\!\Bigl[-4|\alpha_s|^{2}\sin^{2}\!\Bigl(\frac{\omega_s T}{2}\Bigr)\Bigr]\,
  \exp\!\Bigl[-4|\alpha_{\mathrm{LO}}|^{2}\sin^{2}\!\Bigl(\frac{\omega_{\mathrm{LO}}' T}{2}\Bigr)\Bigr],\\
\langle H'\rangle_{\text{cc}}
&=\hbar\omega_s\Bigl(|\alpha_s|^2+\tfrac{1}{2}\Bigr)+\hbar\omega_{\mathrm{LO}}'\Bigl(|\alpha_{\mathrm{LO}}|^2+\tfrac{1}{2}\Bigr),\\
(\Delta H')^2_{\text{cc}}
&=\hbar^2\Bigl(\omega_s^2|\alpha_s|^2+(\omega_{\mathrm{LO}}')^2|\alpha_{\mathrm{LO}}|^2\Bigr).
\end{align}

\paragraph{Squeezed-vacuum$\otimes$coherent.}
The signal factor remains \eqref{eq:app_F_sv_signal}, the LO factor uses $\omega_{\mathrm{LO}}'$:
\begin{equation}
F_{\text{sv}\otimes\text{c}}'(T)
=F_{\mathrm{sv}}(T)\;
 \exp\!\Bigl[-4|\alpha_{\mathrm{LO}}|^{2}\sin^{2}\!\Bigl(\frac{\omega_{\mathrm{LO}}' T}{2}\Bigr)\Bigr],
\end{equation}
with
\begin{equation}
(\Delta H')^2_{\text{sv}\otimes\text{c}}
=\hbar^2\Bigl(\omega_s^2\,2\sinh^2 r\,\cosh^2 r+(\omega_{\mathrm{LO}}')^2|\alpha_{\mathrm{LO}}|^2\Bigr).
\end{equation}

\paragraph{Squeezed-coherent$\otimes$coherent.}
The signal fidelity is \eqref{eq:app_F_sc_signal_general}, and the LO factor uses $\omega_{\mathrm{LO}}'$:
\begin{equation}
F_{\text{sc}\otimes\text{c}}'(T)
=F_{\mathrm{sc}}^{(s)}(T)\;
 \exp\!\Bigl[-4|\alpha_{\mathrm{LO}}|^{2}\sin^{2}\!\Bigl(\frac{\omega_{\mathrm{LO}}' T}{2}\Bigr)\Bigr].
\end{equation}
Energy moments follow from Appendix~\ref{app:two-mode-QSL} by $\omega_{\mathrm{LO}}\to\omega_{\mathrm{LO}}'$.

\section{Refined Quantum Time Resolution in Interferometry}\label{app:quantum-time-res}

Consider two modes at frequencies $\omega_s$ and $\omega_{\mathrm{LO}}$ combined on a balanced beam splitter and measured by difference photocurrent. The operator for the output intensity difference is
\begin{equation}
\hat I_{\mathrm{diff}}
=\hat a_s^\dagger \hat a_{\mathrm{LO}}+\hat a_{\mathrm{LO}}^\dagger \hat a_s,
\label{eq:app_Idiff_def}
\end{equation}
consistent with the standard beam-splitter relations \cite{MandelWolf1995}.

Write the relative optical phase as $\delta\psi(t)$. In a heterodyne setting with a constant detuning $\Delta\omega=\omega_s-\omega_{\mathrm{LO}}$, one has $\delta\psi(t)=\Delta\omega\,t+\delta\psi_0$. For general phase evolution $\delta\psi(t)$, error propagation gives
\begin{equation}
\Delta\psi
=\frac{\Delta I_{\mathrm{diff}}}{\bigl|\partial_{\delta\psi}\langle \hat I_{\mathrm{diff}}\rangle\bigr|},
\qquad
\Delta t
=\frac{\Delta\psi}{\bigl|d\delta\psi/dt\bigr|}.
\label{eq:app_time_resolution_general}
\end{equation}
For constant detuning, $|d\delta\psi/dt|=|\Delta\omega|$, hence $\Delta t=\Delta\psi/|\Delta\omega|$. This local linear phase model is the one used in Secs.~\ref{sec:time-resolution} and \ref{sec:applications_main} when translating between beat-phase accumulation and time resolution.

A quantum Cram\'er-Rao formulation yields, for unbiased local estimators, a state-dependent lower bound
\begin{equation}
\Delta t
\ge \frac{1}{\sqrt{\mathcal{F}_Q}}\;\frac{1}{\bigl|d\delta\psi/dt\bigr|},
\label{eq:app_QCR_time}
\end{equation}
where $\mathcal{F}_Q$ is the quantum Fisher information associated with $\delta\psi$ in the chosen probe family \cite{Helstrom1976,BraunsteinCaves1994}.
Appendices~\ref{app:coh-coh-BHD} and \ref{app:sqcoh-BHD} give explicit $\langle \hat I_{\mathrm{diff}}\rangle$ and $\Delta I_{\mathrm{diff}}$ for coherent and squeezed-coherent inputs.

\section{Balanced detection with coherent-coherent input}\label{app:coh-coh-BHD}

Let the input be a product of coherent states,
\begin{equation}
\ket{\Psi}=\ket{\alpha_s e^{i\psi_s}} \otimes \ket{\alpha_{\mathrm{LO}} e^{i\psi_{\mathrm{LO}}}}.
\end{equation}
A 50:50 beam splitter implements
\begin{equation}
\hat a_{\mathrm{out1}}=\frac{\hat a_s+\hat a_{\mathrm{LO}}}{\sqrt2},\qquad
\hat a_{\mathrm{out2}}=\frac{\hat a_s-\hat a_{\mathrm{LO}}}{\sqrt2}.
\end{equation}
The difference intensity operator reduces to
\begin{equation}
\hat I_{\mathrm{diff}}
=\hat a_{\mathrm{out1}}^\dagger \hat a_{\mathrm{out1}}-\hat a_{\mathrm{out2}}^\dagger \hat a_{\mathrm{out2}}
=\hat a_s^\dagger \hat a_{\mathrm{LO}}+\hat a_{\mathrm{LO}}^\dagger \hat a_s,
\end{equation}
matching \eqref{eq:app_Idiff_def}.

Using factorization on the product state and $\langle \hat a\rangle=\alpha$ for coherent states,
\begin{equation}
\langle \hat I_{\mathrm{diff}}\rangle
=\langle \hat a_s^\dagger\rangle\langle \hat a_{\mathrm{LO}}\rangle
+\langle \hat a_{\mathrm{LO}}^\dagger\rangle\langle \hat a_s\rangle
=\alpha_s^*\alpha_{\mathrm{LO}}\,e^{-i(\psi_s-\psi_{\mathrm{LO}})}
+\alpha_s\alpha_{\mathrm{LO}}^*\,e^{i(\psi_s-\psi_{\mathrm{LO}})}.
\end{equation}
Writing $\delta\psi=\psi_s-\psi_{\mathrm{LO}}$ and taking $\alpha_s,\alpha_{\mathrm{LO}}\ge 0$ gives
\begin{equation}
\langle \hat I_{\mathrm{diff}}\rangle=2\alpha_s\alpha_{\mathrm{LO}}\cos(\delta\psi).
\label{eq:app_Idiff_mean_cc}
\end{equation}

Expand
\begin{equation}
\hat I_{\mathrm{diff}}^2
=(\hat a_s^\dagger \hat a_{\mathrm{LO}})^2+(\hat a_{\mathrm{LO}}^\dagger \hat a_s)^2
+\hat a_s^\dagger \hat a_{\mathrm{LO}}\hat a_{\mathrm{LO}}^\dagger \hat a_s
+\hat a_{\mathrm{LO}}^\dagger \hat a_s \hat a_s^\dagger \hat a_{\mathrm{LO}}.
\end{equation}
On a product of coherent states,
$\langle \hat a^2\rangle=\alpha^2$,
$\langle \hat a^{\dagger 2}\rangle=\alpha^{*2}$,
$\langle \hat a^\dagger \hat a\rangle=|\alpha|^2$,
$\langle \hat a\hat a^\dagger\rangle=|\alpha|^2+1$.
Thus
\begin{align}
\bigl\langle(\hat a_s^\dagger \hat a_{\mathrm{LO}})^2\bigr\rangle
&=\langle \hat a_s^{\dagger 2}\rangle\langle \hat a_{\mathrm{LO}}^{2}\rangle
=\alpha_s^{*2}\alpha_{\mathrm{LO}}^{2},\\
\bigl\langle(\hat a_{\mathrm{LO}}^\dagger \hat a_s)^2\bigr\rangle
&=\langle \hat a_{\mathrm{LO}}^{\dagger 2}\rangle\langle \hat a_s^{2}\rangle
=\alpha_{\mathrm{LO}}^{*2}\alpha_s^{2},\\
\langle \hat a_s^\dagger \hat a_{\mathrm{LO}}\hat a_{\mathrm{LO}}^\dagger \hat a_s\rangle
&=\langle \hat a_s^\dagger \hat a_s\rangle\langle \hat a_{\mathrm{LO}}\hat a_{\mathrm{LO}}^\dagger\rangle
=|\alpha_s|^2\,(|\alpha_{\mathrm{LO}}|^2+1),\\
\langle \hat a_{\mathrm{LO}}^\dagger \hat a_s \hat a_s^\dagger \hat a_{\mathrm{LO}}\rangle
&=\langle \hat a_{\mathrm{LO}}^\dagger \hat a_{\mathrm{LO}}\rangle\langle \hat a_s \hat a_s^\dagger\rangle
=|\alpha_{\mathrm{LO}}|^2\,(|\alpha_s|^2+1).
\end{align}
Hence
\begin{equation}
\langle \hat I_{\mathrm{diff}}^2\rangle
=\alpha_s^{*2}\alpha_{\mathrm{LO}}^{2}+\alpha_{\mathrm{LO}}^{*2}\alpha_s^{2}
+2|\alpha_s|^2|\alpha_{\mathrm{LO}}|^2+|\alpha_s|^2+|\alpha_{\mathrm{LO}}|^2.
\end{equation}
From \eqref{eq:app_Idiff_mean_cc},
$\langle \hat I_{\mathrm{diff}}\rangle^2
=\alpha_s^{*2}\alpha_{\mathrm{LO}}^{2}+\alpha_{\mathrm{LO}}^{*2}\alpha_s^{2}
+2|\alpha_s|^2|\alpha_{\mathrm{LO}}|^2$,
so the variance is
\begin{equation}
\Delta I_{\mathrm{diff}}^2
=\langle \hat I_{\mathrm{diff}}^2\rangle-\langle \hat I_{\mathrm{diff}}\rangle^2
=|\alpha_s|^2+|\alpha_{\mathrm{LO}}|^2.
\label{eq:app_Idiff_var_cc}
\end{equation}

Error propagation yields
\begin{equation}
\Delta\psi
=\frac{\Delta I_{\mathrm{diff}}}{\bigl|\partial_{\delta\psi}\langle \hat I_{\mathrm{diff}}\rangle\bigr|}
=\frac{\sqrt{|\alpha_s|^2+|\alpha_{\mathrm{LO}}|^2}}
{2|\alpha_s|\,|\alpha_{\mathrm{LO}}|\,|\sin(\delta\psi)|}.
\end{equation}
The smallest phase uncertainty is obtained near $ |\sin(\delta\psi)|\approx 1 $, corresponding to the usual linearized phase-sensitive operating point. If $\delta\psi(t)=(\omega_s-\omega_{\mathrm{LO}})t$, then
\begin{equation}
\Delta t
=\frac{\Delta\psi}{|\omega_s-\omega_{\mathrm{LO}}|}
=\frac{\sqrt{|\alpha_s|^2+|\alpha_{\mathrm{LO}}|^2}}
{2|\omega_s-\omega_{\mathrm{LO}}|\,|\alpha_s|\,|\alpha_{\mathrm{LO}}|\,|\sin((\omega_s-\omega_{\mathrm{LO}})t)|}.
\end{equation}

\section{Balanced detection with squeezed-coherent signal and coherent LO}\label{app:sqcoh-BHD}

Let the signal be a squeezed-coherent state in the displaced squeezed vacuum convention,
\begin{equation}
\ket{\alpha_s,\zeta}=\hat D(\alpha_s)\hat S(\zeta)\ket{0},
\qquad
\zeta=r\,e^{i\theta},
\end{equation}
and let the LO be coherent $\ket{\alpha_{\mathrm{LO}}e^{i\psi_{\mathrm{LO}}}}$, with $\alpha_{\mathrm{LO}}\ge 0$ denoting the real nonnegative LO amplitude and $\psi_{\mathrm{LO}}$ its optical phase. The total input is $\ket{\Psi}=\ket{\alpha_s,\zeta}\otimes\ket{\alpha_{\mathrm{LO}}e^{i\psi_{\mathrm{LO}}}}$.
The measured operator is $\hat I_{\mathrm{diff}}=\hat a_s^\dagger \hat a_{\mathrm{LO}}+\hat a_{\mathrm{LO}}^\dagger \hat a_s$.

We Define
\begin{equation}
N=\sinh^2 r,\qquad
M=-e^{i\theta}\sinh r\,\cosh r=-\frac{1}{2} e^{i\theta}\sinh(2r).
\end{equation}
For $\ket{\alpha_s,\zeta}$,
\begin{equation}
\langle \hat a_s\rangle=\alpha_s,\qquad
\langle \hat a_s^2\rangle=\alpha_s^2+M,\qquad
\langle \hat a_s^{\dagger 2}\rangle=\alpha_s^{*2}+M^*,\qquad
\langle \hat n_s\rangle=|\alpha_s|^2+N.
\label{eq:app_sqcoh_moments}
\end{equation}

Using LO factorization $\langle \hat a_{\mathrm{LO}}\rangle=\alpha_{\mathrm{LO}}e^{i\psi_{\mathrm{LO}}}$,
\begin{equation}
\langle \hat I_{\mathrm{diff}}\rangle
=\alpha_{\mathrm{LO}}e^{i\psi_{\mathrm{LO}}}\,\alpha_s^*
+\alpha_{\mathrm{LO}}e^{-i\psi_{\mathrm{LO}}}\,\alpha_s.
\end{equation}
Writing $\alpha_s=|\alpha_s|e^{i\psi_s}$ gives
\begin{equation}
\langle \hat I_{\mathrm{diff}}\rangle
=2|\alpha_s|\,\alpha_{\mathrm{LO}}\cos(\delta\psi),
\qquad \delta\psi=\psi_s-\psi_{\mathrm{LO}}.
\label{eq:app_Idiff_mean_sqcoh}
\end{equation}

Expanding $\hat I_{\mathrm{diff}}^2$ as in Appendix~\ref{app:coh-coh-BHD} and using
$\langle \hat a_{\mathrm{LO}}^2\rangle=\alpha_{\mathrm{LO}}^2 e^{i2\psi_{\mathrm{LO}}}$,
$\langle \hat a_{\mathrm{LO}}^{\dagger 2}\rangle=\alpha_{\mathrm{LO}}^2 e^{-i2\psi_{\mathrm{LO}}}$,
$\langle \hat a_{\mathrm{LO}}\hat a_{\mathrm{LO}}^\dagger\rangle=\alpha_{\mathrm{LO}}^2+1$,
$\langle \hat a_{\mathrm{LO}}^\dagger\hat a_{\mathrm{LO}}\rangle=\alpha_{\mathrm{LO}}^2$,
together with \eqref{eq:app_sqcoh_moments}, yields
\begin{equation}
\langle \hat I_{\mathrm{diff}}^2\rangle
=\alpha_{\mathrm{LO}}^2 e^{i2\psi_{\mathrm{LO}}}\,(\alpha_s^{*2}+M^*)
+\alpha_{\mathrm{LO}}^2 e^{-i2\psi_{\mathrm{LO}}}\,(\alpha_s^{2}+M)
+(2\alpha_{\mathrm{LO}}^2+1)\,(\,|\alpha_s|^2+N\,)+\alpha_{\mathrm{LO}}^2.
\end{equation}
From \eqref{eq:app_Idiff_mean_sqcoh},
$\langle \hat I_{\mathrm{diff}}\rangle^2
=\alpha_{\mathrm{LO}}^2 e^{i2\psi_{\mathrm{LO}}}\alpha_s^{*2}
+\alpha_{\mathrm{LO}}^2 e^{-i2\psi_{\mathrm{LO}}}\alpha_s^{2}
+2\alpha_{\mathrm{LO}}^2|\alpha_s|^2$,
so the variance becomes
\begin{equation}
\Delta I_{\mathrm{diff}}^2
=|\alpha_s|^2+\alpha_{\mathrm{LO}}^2+(2\alpha_{\mathrm{LO}}^2+1)N
+2\,\mathrm{Re}\!\bigl(\alpha_{\mathrm{LO}}^2 e^{-i2\psi_{\mathrm{LO}}}M\bigr).
\label{eq:app_Idiff_var_sqcoh_general}
\end{equation}
Choosing the LO phase to align with the squeezed quadrature, $2\psi_{\mathrm{LO}}=\theta$, gives
\begin{equation}
2\,\mathrm{Re}\!\bigl(\alpha_{\mathrm{LO}}^2 e^{-i2\psi_{\mathrm{LO}}}M\bigr)
=-\alpha_{\mathrm{LO}}^2\sinh(2r),
\end{equation}
and therefore
\begin{equation}
\Delta I_{\mathrm{diff}}^2
=|\alpha_s|^2+\alpha_{\mathrm{LO}}^2 e^{-2r}+\sinh^2 r.
\label{eq:app_Idiff_var_sqcoh_aligned}
\end{equation}

From \eqref{eq:app_Idiff_mean_sqcoh},
\begin{equation}
\bigl|\partial_{\delta\psi}\langle \hat I_{\mathrm{diff}}\rangle\bigr|
=2|\alpha_s|\,\alpha_{\mathrm{LO}}|\sin(\delta\psi)|.
\end{equation}
Thus,
\begin{equation}
\Delta\psi
=\frac{\Delta I_{\mathrm{diff}}}{2|\alpha_s|\,\alpha_{\mathrm{LO}}|\sin(\delta\psi)|},
\end{equation}
with $\Delta I_{\mathrm{diff}}=\sqrt{\Delta I_{\mathrm{diff}}^2}$ from \eqref{eq:app_Idiff_var_sqcoh_general} or \eqref{eq:app_Idiff_var_sqcoh_aligned}. The smallest phase uncertainty is obtained near $ |\sin(\delta\psi)|\approx 1 $, corresponding to the usual linearized phase-sensitive operating point. If $\delta\psi(t)=(\omega_s-\omega_{\mathrm{LO}})t$, then $\Delta t=\Delta\psi/|\omega_s-\omega_{\mathrm{LO}}|$ as in \eqref{eq:app_time_resolution_general}.

\section{Rectangular-window phase averaging}
\label{F}

For an untracked, open-loop integration window during transient acquisition prior
to active phase-locked loop (PLL) engagement, we model the stochastic part of the
demodulated, unwrapped, linearized phase record as a zero-mean, locally stationary
noise process \(n_\psi(t)\). We use the one-sided phase-noise PSD convention in
ordinary frequency \(f\), so that \(S_\psi(f)\) has units
\(\mathrm{rad}^2/\mathrm{Hz}\) and
\begin{equation}
\langle n_\psi(t)n_\psi(t+\tau)\rangle
=
\int_0^\infty S_\psi(f)\cos(2\pi f\tau)\,df .
\label{eq:one_sided_phase_psd_convention_main}
\end{equation}
For a rectangular window average,
\begin{equation}
\hat n_{\psi,T}
=
\frac{1}{T}\int_0^T n_\psi(t)\,dt ,
\label{eq:rectangular_phase_average_main}
\end{equation}
the corresponding transfer function is
\begin{equation}
H_T(f)
=
\frac{1}{T}\int_0^T e^{-i2\pi ft}\,dt
=
e^{-i\pi fT}\frac{\sin(\pi fT)}{\pi fT}.
\label{eq:rectangular_average_filter_main}
\end{equation}
Hence the variance of the averaged stochastic phase is
\begin{equation}
\mathrm{Var}(\hat n_{\psi,T})
=
\int_0^\infty S_\psi(f)|H_T(f)|^2\,df .
\label{eq:phase_average_variance_general_main}
\end{equation}
If \(S_\psi(f)\) is approximately constant over the frequencies appreciably
weighted by \(|H_T(f)|^2\), we may write \(S_\psi(f)\simeq S_\psi\), giving
\begin{equation}
\mathrm{Var}(\hat n_{\psi,T})
=
S_\psi\int_0^\infty
\left[\frac{\sin(\pi fT)}{\pi fT}\right]^2df
=
\frac{S_\psi}{2T}.
\label{eq:white_phase_average_variance_main}
\end{equation}

\end{document}